\documentclass{aa}  

\usepackage{graphicx}
\usepackage{txfonts}
\usepackage{xcolor}
\usepackage{subcaption} % necessary for continued figures
\usepackage{lscape}     % to rotate a single page table
\usepackage{placeins}           % useful with \FloatBarrier
\usepackage{comment}
\usepackage{hyperref}
\usepackage{adjustbox}

\begin{document}

   \title{The ALMA carbon monoxide supernova (ACOS) survey}

   % \subtitle{II. Turbulent molecular gas at giant molecular cloud scales at the positions of core-collapse supernovae}
   \subtitle{II. Turbulent giant molecular clouds at the positions of core-collapse supernovae}

   \author{M. Solar\inst{1}\fnmsep\thanks{martin.solar@amu.edu.pl}
        \and M. J. Michałowski\inst{1}\thanks{michal.michalowski@amu.edu.pl}
        \and J. Nadolny\inst{1}\fnmsep\inst{2}\fnmsep\inst{3}
        \and J. Sollerman\inst{4}
        \and E. Zapartas\inst{5}\fnmsep\inst{6}
        \and A. Oliva\inst{7}
        \and L. Galbany\inst{8}\fnmsep\inst{9}
        \and J. Hjorth\inst{10}
        \and B. Ayala\inst{11}        
        \and M. Koprowski\inst{12}
        \and A. Leśniewska\inst{1}\fnmsep\inst{10}
        \and P. Nowaczyk\inst{1}
        \and O. Ryzhov\inst{1}
        \and B. \v{S}laus\inst{1}
        }

   \institute{Astronomical Observatory Institute, Faculty of Physics and Astronomy, Adam Mickiewicz University, ul. Słoneczna 36, 60-286, Poznań, Poland
            \and Instituto de Astrofísica de Canarias, E-38205 La Laguna, Tenerife, Spain
            \and Departamento de Astrofísica, Universidad de La Laguna (ULL), E-38205 La Laguna, Tenerife, Spain
            \and The Oskar Klein, Department of Astronomy, Stockholm University, Albanova University Center, Stockholm, Sweden
            \and Institute of Astrophysics, Foundation for Research and Technology-Hellas, 71110 Heraklion, Greece
            \and Physics Department, National and Kapodistrian University of Athens, 15784 Athens, Greece
            \and Space Research Center (CINESPA), School of Physics, University of Costa Rica, 11501 San José, Costa Rica
            \and Institute of Space Sciences (ICE, CSIC), Campus UAB, Barcelona, Spain
            \and Institut d’Estudis Espacials de Catalunya (IEEC), Barcelona, Spain
            \and DARK, Niels Bohr Institute, University of Copenhagen, Jagtvej 155A, DK-2200 Copenhagen N, Denmark
            \and Millennium Institute of Astrophysics, Nuncio Monseñor Sotero Sanz 100, Of. 104, Providencia, Santiago, Chile
            \and Institute of Astronomy, Faculty of Physics, Astronomy and Informatics, Nicolaus Copernicus University, Toruń, Poland
            }

   \date{Received Month Day, Year}

  \abstract
  % context heading (optional)
   {Study of cold molecular hydrogen gas (hereafter molecular gas) provides crucial insights into its interplay with star-forming regions. % where stars are born.
    However, the connection between molecular gas turbulence and the sites of massive star ($\gtrsim 8$--$10$\,M$_{\sun}$) explosions as core-collapse supernovae (CCSNe) remains unexplored.
    }
  % aims heading (mandatory)
   {We measure for the first time the turbulence of molecular gas in environments of CCSNe, with the aim to constrain the nature of their progenitors.
    } 
  % methods heading (mandatory)
   {In order to reach spatial resolutions of giant molecular cloud (GMC) sizes ($\sim$\,$100$\,pc), we collected ALMA carbon monoxide $J = 2 \rightarrow 1$ spectral line ($\sim$\,$230.54\,\mathrm{GHz}$) observations (as a tracer of molecular gas) at the locations of 33 nearby CCSNe ($< 100$\,Mpc).
    }
  % results heading (mandatory)
   {
   % We found that CCSNe prefer regions of highly turbulent GMCs compared to the average content of their host galaxies.
   We found that CCSNe prefer molecular gas regions with high velocity dispersion compared to the average of their host galaxies.}
  % conclusions heading (optional)
   {For CCSN progenitors, this observational evidence supports their increased formation in regions of high densities and/or their binary nature.}

   \maketitle
   \nolinenumbers

%%%%%%%%%%%%%%%%%%%%%%%%%%%%%%%%%%%%%%%%%%%%%%%%%%%%%%%%%%%%%%
\section{Introduction} \label{sec:sec1}
Supernovae (SNe) play a crucial role in galaxy evolution by driving positive feedback and turbulent pressure that creates over-dense regions, necessary to form stars \citep{1977McKee}.
Core-collapse supernovae (CCSNe) occur in the final stages of massive star evolution \citep[with initial masses M$_{\rm{init}} \gtrsim 8$--$10$\,M$_{\sun}$;][and references therein]{2009bSmartt}, after the iron core collapses under gravity.
However, the observed explosions exhibit different properties, such as the amount of energy released, velocity of the expansion, element production, and level of interaction with the circumstellar medium.
In general, the primary physical characteristics of CCSNe, which directly influence how these explosions are classified, depend on the progenitor properties such as initial mass, metallicity, rotation, mass loss, and binary interaction \citep[][and references therein]{2012Langer}.

CCSNe are mainly classified according to their observed spectroscopic features in the visible range, such as the presence or absence of some characteristic lines \citep[][and references therein]{1997Filippenko}.
Typically, these features are measured at the peak of the optical lightcurve --- within the first couple of weeks after explosion --- when the ejecta remain optically thick.
Type II (H-rich) SNe exhibit prominent hydrogen lines which persist throughout the entire photospheric phase \citep[$\approx 50$--$100$ days after explosion;][]{1981Branch,2001Leonard}.
In contrast, stripped–envelope SNe (SESNe)
include Type IIb showing hydrogen which disappears a few weeks after the explosion when helium lines emerge \citep{1993Filippenko}, Type Ib missing hydrogen but displaying helium lines \citep{1987Harkness}, and Type Ic lacking both hydrogen and helium lines \citep{1986Wheeler}.

SESNe, without detections of hydrogen/helium lines, are explained as originating from progenitor stars where the envelope was removed prior to the explosion.
The progenitors of SESNe are believed to come from two different channels.
The classical interpretation involves very massive stars (M$_{\rm{init}}$\,$\approx$\,20--30\,M$_{\sun}$), from which the star passes through a Wolf-Rayet phase and loses their outer layers due to strong stellar winds \citep{1993Woosley}.
They are short-lived with lifetimes less than $5$\,Myr.
The other scenario is that the external layers of SESN progenitors are stripped by mass transfer due to binary interaction \citep{1992aPodsiadlowski,2010Yoon,2013Eldridge,2021Laplace,2025Gilkis,2026Souropanis}.
The mass transfer mechanism is efficient for progenitors at lower masses (M$_{\rm{init}}< 20$\,M$_{\sun}$), consequently living longer than in the single massive star counterpart scenario.

Around $70\%$ of the Galactic O-type stars are expected to interact with a companion \citep{2012Sana}.
\cite{2025Villasenor} found a multiplicity fraction (the fraction of stars that have companions) of $\sim$\,$80\%$ for early B-type stars (with initial masses between 8 and 15 M$_{\odot}$).
From population synthesis simulations, H-rich SN progenitors have a significant fraction (from $1/3$ to $1/2$) of mergers or binary history (i.e., to have exchanged material, usually gaining mass; \citealt{1992aPodsiadlowski,2019Zapartas,2021bZapartas,2024Schneider}). 
This is also supported with observational evidence of possible binary history in exotic evens \citep[e. g., \object{SN1987A};][]{1992bPodsiadlowski,2017Menon}, as well as even in more normal-looking Type II SNe \citep{2023Bostroem,2026Niu}.
In addition, the multiplicity fraction is expected to increase with the stellar initial mass \citep[][and references therein]{2023Offner}.
However, it is not clear how the binary parameters (such as mass ratio, type of interaction, eccentricity, and periodicity) affect their progenitors' evolution and explosion properties \citep[][and references therein]{2024Marchant}.

One approach that allows us to comprehend the SN progenitor properties is the statistical study of their immediate environments \citep[][and references therein]{2015Anderson}.
Specifically, the cold molecular hydrogen (H$_{2}$) gas (hereafter molecular gas) encodes valuable information on the spatial distribution and formation sites of SN progenitors within galaxies.
This is because the molecular gas represents the fuel reservoir from which stars are born.
Previous works already investigated the molecular gas in the environments of SNe \citep{2017Galbany,2023MaykerChen,2024Solar} using carbon monoxide (CO) as a proxy of H$_{2}$ \citep[][and references therein]{2013Bolatto}.
However, the turbulence of the molecular gas at the CCSN positions remains unknown, because no studies have addressed this topic.

Turbulence is defined as the irregular and chaotic motions of particles \citep{1941Kolmogorov}.
In the context of the cold interstellar medium (ISM), velocity dispersion of the molecular gas is used as a proxy of turbulence across various phases and scales \citep[][and references therein]{2004Elmegreen,2025Narayan}.
Theoretical models predict whether a given molecular cloud collapses under self-gravity or disperses due to turbulent motions \citep{1992McKee}.
For example, the supersonic turbulent shocks produce localized density enhancements (filaments or sheets).
This can trigger gravitational collapse (positive feedback), while simultaneously in other regions of the cloud the collapse is hindered by the turbulence \citep[negative feedback;][]{2004MacLow,2015Federrath}.
\cite{2024Zhao} suggested that star formation, feedback and gas properties are connected at different scales (from sub-parsecs to several hundreds of parsecs).
Specifically, in turbulent and dense environments, core fragmentation triggers multiple independent collapses \citep{2004Goodwin,2007Goodwin, 2026Luo}.
This core fragmentation is considered the main driver for forming multiple systems \citep{2026Luo}.
In order to understand the conditions from which the progenitors of CCSNe are born, it is essential to measure the dynamical state of the cold ISM at these explosion sites.

To this aim, we started the Atacama Large Millimeter/submillimeter Array\footnote{https://almascience.eso.org/} (ALMA) CO SN (ACOS) survey (we refer to \citealt{2024Solar} as Paper I of this series).
The goal is to infer the properties of SN progenitors, using high spatial resolutions of the order of typical giant molecular cloud (GMC) sizes \citep[$\sim$\,$100$ pc;][]{2016Leroy}.
This is an important factor because the GMC properties can only be accurately constrained if the resolution at least matches their sizes.
In this Paper, we measure the molecular gas turbulence using the velocity dispersion from the fundamental rotational transition $J = 2 \rightarrow 1$ of $^{12}$C$^{16}$O (hereafter CO(2--1), restframe $\sim$\,$230.54$\,GHz or $\sim$\,$1.3$\,mm) at the locations of CCSNe.

The structure of this paper is as follows:
Section~\ref{sec:sec2} introduces the data and sample selection, and the turbulence parameters used, 
Section~\ref{sec:sec3} presents our results,
Section~\ref{sec:sec4} details the interpretation of results, 
and Section~\ref{sec:sec5} summarizes and concludes our work.

%%%%%%%%%%%%%%%%%%%%%%%%%%%%%%%%%%%%%%%%%%%%%%%%%%%%%%%%%%%%%%
% \section{Data and Sample Selection}\label{sec:sec2}

\section{Method}\label{sec:sec2}

\subsection{Data and sample selection}

We compiled the CCSN sample from various databases \citep{1989Barbon, 1993Tsvetkov, 1998Rutledge, 1999Barbon, 2004Tsetkov, 2012Lennarz, 2017Guillochon}. 
We divided the CCSNe into two categories: H-rich SNe (Type II, IIP, and IIn) and SESNe (Type IIb, Ib, Ib/c, Ic, and Ic-BL).
Thermonuclear SNe (Type Ia) were excluded from the analysis because of their different progenitor evolution properties and explosion mechanisms.

% We gathered CO(2--1) data from the ACOS survey \citep{2024Solar}, the 
% Physics at High Angular resolution in Nearby Galaxies\footnote{https://sites.google.com/view/phangs/home} (PHANGS) survey  \citep{2021aLeroy,2021bLeroy}, and ALMA archival data\footnote{https://almascience.eso.org/aq/}.

We gathered CO(2--1) data from the 
Physics at High Angular resolution in Nearby Galaxies\footnote{https://sites.google.com/view/phangs/home} (PHANGS) survey  \citep{2021aLeroy,2021bLeroy}, ACOS survey \citep{2024Solar}, and ALMA archival data\footnote{https://almascience.eso.org/aq/}.
The PHANGS-ALMA survey consists of 90 nearby ($<$\,$20$\,Mpc), face-on ($i<75^{o}$), star-forming (star formation rate; $\log \left( \rm{SFR} / \,M_{\odot}\,yr^{-1}\right) \sim 0.17$--$7$) galaxies, with stellar masses ($M_{*}$) ranging from $10^{9.03}$\,M$_{\odot}$ to $10^{11.10}$\,M$_{\odot}$, observed at the CO(2--1) line frequency.
81 PHANGS-ALMA targets combined 12 m, 7 m, and total power (TP) arrays.
The rest 9 observations combined 7 m array configuration and 7m+TP (not in our sample).
Typical 12m+7m+TP maps cover $\sim$\,$7\,$arcmin$^{2}$ (124 kpc$^{2}$), with a median angular resolution of $1\farcs3$, channel width of 2.54 km\,s$^{-1}$ and 1$\sigma$ sensitivity of $\sim$\,$6.2$\,mJy beam$^{-1}$.
In terms of physical scales, this corresponds to spatial scales of 100 pc and 1$\sigma$ mass sensitivity of $\sim$\,$2\times10^{4}$\,M$_{\odot}$ per channel.

The ACOS survey (ALMA project ID: 2021.1.00099.S, PI: M. Michałowski) is 
%based in increase the sample size of SN CO(2--1) environments 
aimed at increasing the Type Ic SN sample size observed at CO(2--1)
at physical resolutions similar to that of PHANGS-ALMA.
ACOS consists of typical root mean square of $7.0\,$mJy\,beam$^{-1}$  for a bandwidth of 2.049 km\,s$^{-1}$ (1.6\,MHz).
The observations were centered at the SN position in order to obtain the best signal-to-noise ($S/N$) of the parent molecular cloud.
ACOS project does not require the sensitivity to high values of angular scales as it is not aimed at studying the kiloparsec scales and there is no need to use extra observations with more compact arrays, such as 7 m or TP.
%Same as PHANGS-ALMA observations, the selected lines were CO v=0 $J = 2 \rightarrow 1$, SiO v=1 $J = 5 \rightarrow 4$, and $^{13}$CO v=1 $J = 2 \rightarrow 1$ to cover the maximum of spectral windows.

The observing goals of archival data\footnote{ALMA project IDs: 2015.1.00902.S, 2017.1.00255.S, 2018.1.00272.S, 2018.A.00062.S, 2018.1.00538.S, and 2019.1.01305.S} were different for each of them.
However, our aim was to measure the CO(2--1) emission line reaching typical cloud-scale resolutions.
The angular resolutions ranged from $0\farcs105$ to $1\farcs136$ (four cases) for 12 m arrays and was $5\farcs316$ (one case) for the 7 m array.
% That is why the selected array configuration ranging from $0\farcs105$ to $1\farcs136$ (four cases) for 12 m arrays, and $5\farcs316$ (one case) with the 7 m array.
The resulting spatial resolution for the four cases with 12 m observations is 40--150 pc and for the case of 7 m observations, it is 230 pc.

% The PHANGS team reduced their observations using the version 5.6.1-8 of the Common Astronomy Software Applications \citep[\texttt{CASA};][]{2007McMullin,2022CASATeam}.
% Interferometric and TP data was calibrated, imaged, and deconvoluted using the PHANGS-ALMA pipeline version 4.0 \citep{2021aLeroy}.
% For ACOS and archival data, we used archive file products that perform a standardized calibration, imaging, and deconvolution using \texttt{CASA}.

The PHANGS team reduced their observations using the version 5.6.1-8 of the Common Astronomy Software Applications \citep[\texttt{CASA};][]{2007McMullin,2022CASATeam}.
Interferometric and TP data were calibrated, imaged, and deconvoluted using the PHANGS-ALMA pipeline version 4.0 \citep{2021aLeroy}.
For ACOS and archival data, we used archive file products that were created with a standardized calibration, imaging, and deconvolution using \texttt{CASA}.
The \texttt{CASA} versions used for default data reduction of archive products were the following: 5.1.1-5 (SN2020cuj), 5.4.0-70 (SN2013dk, SN2016adj, and SN2021kwc), 5.6.1-8 (SN2007C, and SN2004ip), and 6.2.1-7 (SN1994ai, SN2005lr, SN2009bb, SN2009gd, SN2011hp, and SN2011jl).
The images for all targets were reconstructed using the \texttt{tclean} task, with a weighting scheme of ``Briggs'' ($weighting=\rm{'}briggs\rm{'}$) and a robustness parameter of 0.5 ($robust=0.5$).
This configuration balances noise and resolution, capturing 95\% of native sensitivity with a highly focused beam.

Only CCSN locations with high $S/N$ CO(2--1) fluxes were considered.
The threshold used is $S/N>4$, the same as the ‘strict' masks from PHANGS-ALMA \citep{2006Rosolowsky, 2021aLeroy}.
A similar method was applied to the PHANGS galaxies, ACOS, and archival data for their detected pixels.
The final sample consists of 33 CCSNe (19 H-rich SNe and 14 SESNe).
21 CCSNe (16 H-rich SNe and 5 SESNe) are from 15 PHANGS-ALMA galaxies and 12 CCSNe (3 H-rich SNe and 9 SESNe) are from 12 galaxies using the ACOS survey and archival data.

%%%%%%%%%%%%%%%%%%%%%%%%%%%%%%%%%%%%%%%%%%%%%%%%%%%%%%%%%%%%%%
% \section{Methods}\label{sec:sec3}
\subsection{Turbulence parameters}

% {\color{blue}Pixel-based integrated intensities do not resolve individually GMCs but cloud-scale observations comprise, on average, their demography \citep[][and references therein]{2024Schinnerer}.
% We refer to cloud-scale observations as tracers of GMC properties \citep{2016Leroy}.}
On average, cloud-scale observations recover GMC properties \citep[][and references therein]{2016Leroy, 2024Schinnerer}.
Although GMC are not individually resolved, we refer to cloud-scale observations as tracers of GMC structures.
From the CO(2--1) line fluxes we computed the properties of molecular gas at GMC scales, such as the molecular gas surface density $\Sigma_{\rm{mol}}$ \citep[corrected by inclination $i$;][]{2014Makarov,2020Lang}.
As our sample comprises of spiral galaxies found on the main sequence of star formation from the Local Universe ($< 100$\,Mpc), we adopted a Milky Way CO(1--0)-to-H$_{2}$ conversion factor (or the mass-to-light ratio) $\alpha_{\rm{CO}}^{1-0} = 4.35$\,M$_{\sun}$\,pc$^{-2}$\,(K\,km\,s$^{-1}$)$^{-1}$ \citep{2013Bolatto} and a CO(2--1)-to-CO(1--0) line ratio R$_{21} = 0.65$ \citep{2021denBrok}.
$\alpha_{\rm{CO}}^{1-0}$ and R$_{21}$ could vary toward galactic centers, resulting in lower \citep{2013Sandstrom} and higher \citep{2020Koba,2021Yajima,2025Koba} values, respectively. 
This is explained by different values of the ISM pressure and gas temperature.
Also, R$_{21}$ depends on the star formation rate \citep{2025denBrok} and the interaction of SN remnants with molecular clouds \citep{2013Jeong,2014Chen,2023Sano}.
We notice that we may overestimate or underestimate $\Sigma_{\rm{mol}}$.

We measured the velocity dispersion of CO(2--1), $\sigma_{\rm{CO}(2-1)}$, from moment 2 maps.
This $\sigma_{\rm{CO}(2-1)}$ is widely used as a proxy of the turbulence in H$_{2}$ \citep{1979Phillips,2025Narayan}. 
From the spatial resolution $\theta$ and the physical observables $\Sigma_{\rm{mol}}$ and $\sigma_{\rm{CO}(2-1)}$, we calculated the dynamical state of the cloud as the virial parameter $\alpha_{\rm{vir}}$ as

\begin{equation}
    \alpha_{\rm{vir}}=\frac{10 \sigma^{2}_{\rm{CO(2-1)}}}{\pi G \Sigma_{\rm{mol}} \theta}.
\end{equation}
This virial parameter quantifies the degree of gravitational boundedness of the cloud \citep{1992Bertoldi, 1992McKee}.
This constrains whether or not molecular gas at GMC scales is in a state of collapse distinguishing a virialized state ($\alpha_{\rm{vir}} < 1$), marginally bound state ($1 < \alpha_{\rm{vir}} < 2$), and dispersing due to turbulent motions ($\alpha_{\rm{vir}} > 2$).
Likewise, it is possible to construct the so-called Heyer-Keto relation \citep{1981Larson, 1986Keto, 2009Heyer} as

\begin{equation}
    \frac{\sigma^{2}_{\rm{CO}(2-1)}}{\theta}=\frac{\pi G \alpha_{\rm{vir}} \Sigma_{\rm{mol}}}{10},
\end{equation}

\noindent which represents a universal law that scales the internal turbulence of molecular gas at GMC scales with their size and surface density.

% The GMC properties for each CCSN site were computed at their coordinate positions.
The molecular gas properties at each CCSN site were computed at their coordinate positions.
The beam size of the observations resolve the cloud scales and with the CCSN positions we extract the environmental properties.
A similar procedure was used for the PHANGS-ALMA sample, but using their detected pixels.
Despite the fact that molecular gas regions are resolved at cloud-scales, in this analysis it is not possible to distinguish between the internal velocity dispersion of individual GMCs and the relative motion of multiple clouds.

Within our sample, PHANGS-ALMA galaxies are completely covered, thus, all their CO-detected pixels are included in the analysis in order to compare their values with the CCSN positions.
In order to have a similar contribution from our PHANGS-ALMA galaxies, we selected randomly 10\,000 pixels (without repetitions) for each of them, resulting in a total of 150\,000 pixels (15 host galaxies).
In that way, statistical calculations are weighted equally for each PHANGS-ALMA galaxy.
% \textbf{Host galaxies.}

%%%%%%%%%%%%%%%%%%%%%%%%%%%%%%%%%%%%%%%%%%%%%%%%%%%%%%%%%%%%%%
\section{Results}\label{sec:sec3}

\begin{figure*}[ht!]
    \centering
    % \resizebox{\hsize}{!}{\includegraphics{fig1.pdf}}
    \includegraphics[width=17.5cm]{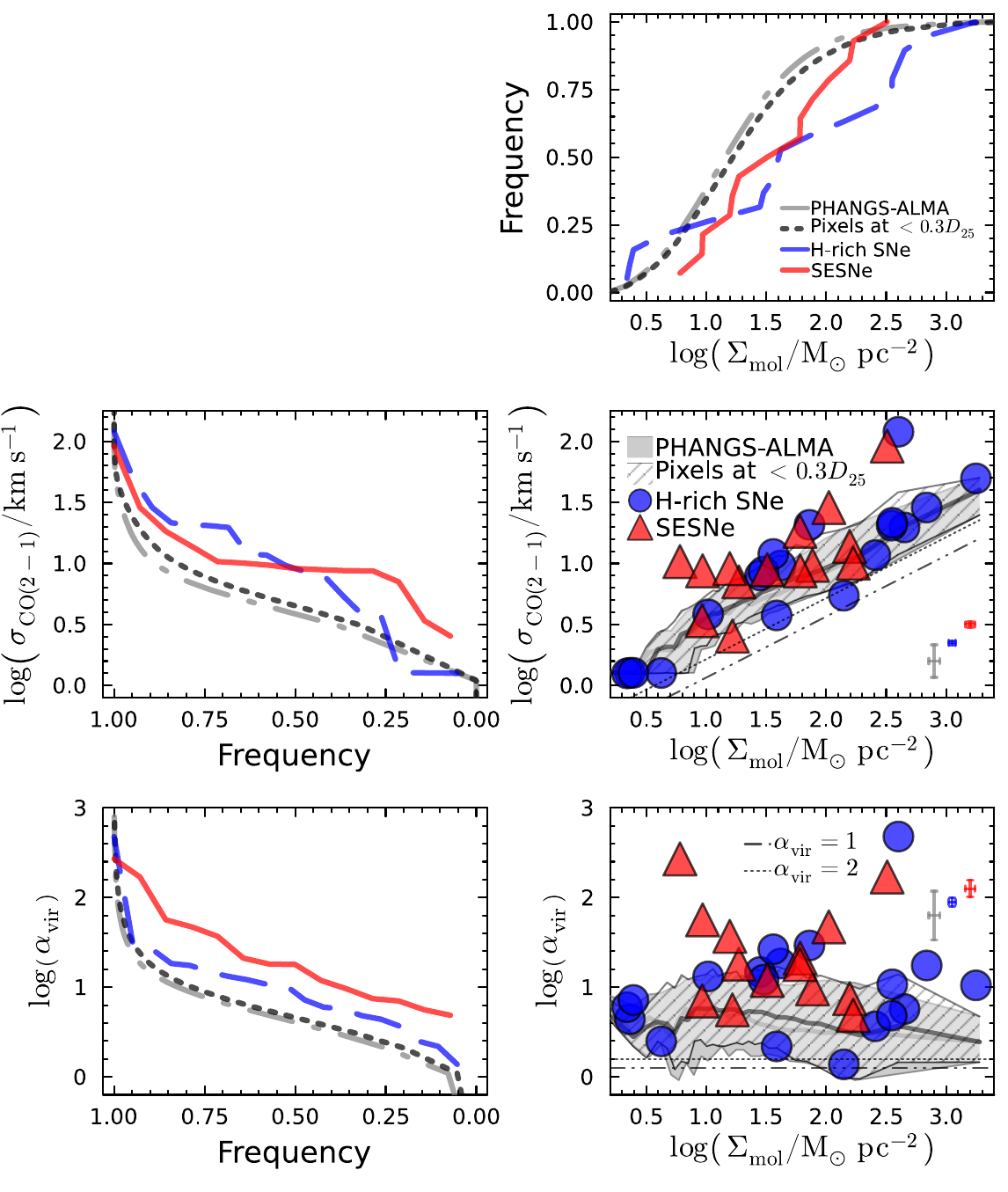}
    \caption{Molecular gas properties for CCSNe and PHANGS-ALMA galaxy pixels:
    CO(2--1) velocity dispersion ($\sigma_{\rm{CO(2-1)}}$; middle right panel) and virial parameter ($\alpha_{\rm{vir}}$; bottom right panel) as a function of the molecular gas surface density ($\Sigma_{\rm{mol}}$), respectively. 
    Locations of H-rich SNe, SESNe, and PHANGS-ALMA galaxy pixels are represented as blue circles, red triangles, and gray shaded regions, respectively.
    Also, pixels at normalized distances lower than 0.3$D_{25}$ are included as hatched shaded regions.
    PHANGS-ALMA galaxy shaded and hatched regions denote their 1$\sigma$ scatter. 
    Typical error bars are denoted as gray, blue, and red for H-rich SNe, SESNe, and PHANGS-ALMA galaxies, respectively.
    The black dashed double-dotted and black dotted lines represent the virial parameter at $\alpha_{\rm{vir}}=1$ and $\alpha_{\rm{vir}}=2$ (from the Heyer-Keto relation using $\theta=100$ pc), respectively.
    The empirical cumulative distribution functions for the positions of $\Sigma_{\rm{mol}}$ (top right panel), $\sigma_{\rm{CO(2-1)}}$ (middle left panel), and $\alpha_{\rm{vir}}$ (bottom left panel) are also included. 
    H-rich SNe, SESNe, and PHANGS-ALMA host galaxy pixel distributions are represented as a blue dashed line, red solid line, and gray dot dashed line, respectively. PHANGS galaxy pixels at distances lower than $0.3D_{25}$ are represented by a black dotted line.}
    \label{fig:fig1}
\end{figure*}

Figure~\ref{fig:fig1} shows the relation of $\sigma_{\rm{CO(2-1)}}$ and $\alpha_{\rm{vir}}$ as a function of $\Sigma_{\rm{mol}}$ for the CCSN sample and random pixels in the host galaxies.
In addition, we show the Heyer--Keto relation for $\alpha_{\rm{vir}} = 1$ and $\alpha_{\rm{vir}} = 2$ using $\theta = 100$ pc.
We include the empirical cumulative distribution functions (eCDFs) for visualization purposes.
Table~\ref{tab:tab1} shows the median values with $1\sigma$ (16\% and 84\%) confidence intervals from 10$^{4}$ Monte Carlo simulations \citep{1999Newman}.
For the purpose of comparison of two different distributions (e.g., H-rich SNe versus SESNe) and assessing whether there is a statistically significant difference (p-value\,$<0.05$), the two-sample tests of Kolmogorov--Smirnov \citep[KS;][]{1951Massey} and Anderson--Darling \citep[AD;][]{1952Anderson} are summarized in Table~\ref{tab:tab2}.
The individual CCSN properties and environments are listed in Table~\ref{tab:tabA1} and Table~\ref{tab:tabA2} (Appendix~\ref{app:appA1}), respectively.

\begin{table*}[ht!]
    \caption{Medians and 1$\sigma$ confidence intervals for each sample of molecular gas density, line width, and virial parameter}
    \label{tab:tab1}
    \centering
    \begin{tabular}{l|ccc}
        \hline
        \hline
        Parameter & $\Sigma_{\rm{mol}}$ & $\sigma_{\rm{CO(2-1)}}$ & $\alpha_{\rm{vir}}$\\
        {} & (M$_{\sun}$ pc$^{-2}$) & (km s$^{-1}$) & {}\\
        \hline
        PHANGS-ALMA galaxies & $14.48^{+0.02}_{-0.02}$ & $3.81^{+0.01}_{-0.01}$ & $3.56^{+0.01}_{-0.01}$ \\
        H-rich SNe           & $42.8^{+8.3}_{-2.0}$    & $9.0^{+2.0}_{-0.6}$    & $9.5^{+1.3}_{-2.4}$ \\
        SESNe                & $44.2^{+3.1}_{-12.4}$   & $9.5^{+0.3}_{-0.3}$    & $16.1^{+3.3}_{-2.3}$ \\
        \hline
    \end{tabular}
\end{table*}

\begin{table*}[ht!]
    \caption{KS and AD two-sample tests applied between the distributions of PHANGS-ALMA galaxy pixels, H-rich SNe, and SESNe. 
    Values of the molecular gas surface density, CO(2--1) velocity dispersion, and virial parameter are compared.
    p-values less than 0.05 are marked as bold}
    \label{tab:tab2}
    \centering
    \begin{tabular}{l|cccc}
        \hline
        \hline
        Two-sample test & KS statistics & KS p-value & AD statistics & AD p-value\\
        \hline
        $\Sigma_{\rm{mol}}$ \,\,\,\,\,\,\,\, $\rightarrow$ PHANGS-ALMA galaxies vs. H-rich SNe & $4.40 \times 10^{-1}$ & $\mathbf{1.45 \times 10^{-3}}$                                                     & $8.97$ & $\mathbf{3.81 \times 10^{-5}}$ \\ %
        $\Sigma_{\rm{mol}}$ \,\,\,\,\,\,\,\, $\rightarrow$ PHANGS-ALMA galaxies vs. SESNe & $3.55 \times 10^{-1}$ & $5.84 \times 10^{-2}$                                                                    & $3.63$ & $\mathbf{1.36 \times 10^{-2}}$\\ %
        $\Sigma_{\rm{mol}}$ \,\,\,\,\,\,\,\, $\rightarrow$ H-rich SNe \,\,\,\,\,\,\,\,\,\,\,\,\,\,\,\,\,\,\,\,\,\,\,\,\,\,\,\,\,\,\,\,\,\,\,\, vs. SESNe & $3.15 \times 10^{-1}$ & $3.98 \times 10^{-1}$     & $7.21$ & $2.18 \times 10^{-1}$ \\
        $\sigma_{\rm{CO}(2-1)}$ $\rightarrow$ PHANGS-ALMA galaxies vs. H-rich SNe & $4.63 \times 10^{-1}$ & $\mathbf{5.83 \times 10^{-4}}$                                                                  & $7.66$ & $\mathbf{4.73 \times 10^{-5}}$ \\ %
        $\sigma_{\rm{CO}(2-1)}$ $\rightarrow$ PHANGS-ALMA galaxies vs. SESNe & $6.48 \times 10^{-1}$ & $\mathbf{1.57 \times 10^{-5}}$                                                                        & $10.74$ & $\mathbf{5.84 \times 10^{-6}}$ \\ %
        $\sigma_{\rm{CO}(2-1)}$ $\rightarrow$ H-rich SNe \,\,\,\,\,\,\,\,\,\,\,\,\,\,\,\,\,\,\,\,\,\,\,\,\,\,\,\,\,\,\,\,\,\,\,\, vs. SESNe & $2.60 \times 10^{-1}$ & $6.50 \times 10^{-1}$                  & $7.21$ & $4.25 \times 10^{-1}$ \\
        $\alpha_{\rm{vir}}$ \,\,\,\,\,\,\,\,\,\, $\rightarrow$ PHANGS-ALMA galaxies vs. H-rich SNe & $3.60 \times 10^{-1}$ & $\mathbf{1.50 \times 10^{-2}}$                                                & $5.17$ & $\mathbf{2.44 \times 10^{-3}}$\\ %
        $\alpha_{\rm{vir}}$ \,\,\,\,\,\,\,\,\,\, $\rightarrow$ PHANGS-ALMA galaxies vs. SESNe & $5.72 \times 10^{-1}$ & $\mathbf{2.01 \times 10^{-4}}$                                                       & $13.67$ & $\mathbf{2.50 \times 10^{-7}}$ \\ %
        $\alpha_{\rm{vir}}$ \,\,\,\,\,\,\,\,\,\, $\rightarrow$ H-rich SNe \,\,\,\,\,\,\,\,\,\,\,\,\,\,\,\,\,\,\,\,\,\,\,\,\,\,\,\,\,\,\,\,\,\,\,\, vs. SESNe & $3.61 \times 10^{-1}$ & $2.45 \times 10^{-1}$ & $7.21$ & $6.64 \times 10^{-2}$ \\
        \hline
    \end{tabular}
\end{table*}

%From Type II SNe and SESNe molecular gas surface density distributions, we prove that they explode in similar conditions.
H-rich SNe and SESNe exhibit similar molecular gas surface density distributions.
Indeed, the median values are $42.8^{+8.3}_{-2.0}$\,M$_{\sun}$\,pc$^{-2}$ and $44.2^{+3.1}_{-12.4}$\,M$_{\sun}$\,pc$^{-2}$ for H-rich SNe and SESNe, respectively.
Furthermore, the KS and AD tests results in high p-values of $3.98\times10^{-1}$ and $2.18\times10^{-1}$, respectively.
On the other hand, both CCSN populations are located in higher molecular gas surface density environments than the average pixels from the PHANGS-ALMA galaxies ($\Sigma_{\rm{mol}}=14.48^{+0.02}_{-0.02}$\,M$_{\sun}$\,pc$^{-2}$).
The clear offset of H-rich SNe and PHANGS-ALMA galaxies is represented by low p-values of the KS and AD tests of $1.50\times10^{-3}$ and $3.81\times10^{-5}$, respectively.
The difference between SESNe and PHANGS-ALMA galaxy pixels is not as significant with p-values for KS of $5.84\times10^{-2}$ but for AD of $1.36\times10^{-2}$.

A similar analysis is repeated for the velocity dispersion of CO(2--1) for both types of CCSNe.
Their medians are $9.0^{+2.0}_{-0.6}$\,km\,s$^{-1}$ and $9.5^{+0.3}_{-0.3}$\,km\,s$^{-1}$ for H-rich SNe and SESNe, respectively.
We thus observe a similar velocity dispersion for SESNe as for H-rich SNe. 
Their p-values of the KS and AD tests are $6.50\times10^{-1}$ and $4.25\times10^{-1}$, respectively. 
On the other hand, the velocity dispersions for both CCSN populations are more extreme than the average content of the PHANGS-ALMA galaxy pixels ($3.81^{+0.01}_{-0.01}$\,km\,s$^{-1}$).
This is also evident from the low p-values of the KS (AD) test of $5.83 \times 10^{-4}$ ($4.73 \times 10^{-5}$) and $1.57 \times 10^{-5}$ ($5.84 \times 10^{-6}$) comparing PHANGS-ALMA galaxy pixels and H-rich SNe and SESNe, respectively.

The last property analyzed here is the virial parameter.
SESNe ($\alpha_{\rm{vir}}=16.1^{+3.3}_{-2.3}$) are located in molecular gas regions with higher velocity dispersion than H-rich SNe ($\alpha_{\rm{vir}}=9.5^{+1.3}_{-2.4}$).
However, this is not statistically significant, as the comparison of the two distributions gives the p-values of $2.45\times10^{-1}$ and $6.64\times10^{-2}$ for the KS and AD tests, respectively.
On the other hand, $\alpha_{\rm{vir}}$ for PHANGS-ALMA pixels is $3.56^{+0.01}_{-0.01}$.
Comparing the distributions of H-rich SNe and PHANGS-ALMA pixels, we obtain p-values of $1.45\times10^{-2}$ and $2.44\times10^{-3}$ for the KS and AD tests, respectively.
It is clear that both samples come from different distributions and there is a trend for H-rich SNe occurring in molecular gas with higher velocity dispersion than seen in the PHANGS-ALMA pixels.
A similar pattern occurs for PHANGS-ALMA pixels and SESNe, with the p-values of $2.01\times10^{-4}$ and $2.50\times10^{-7}$ for the KS and AD tests, respectively.
The result is statistically significant, indicating evidence against the null hypothesis that the two samples come from the same distribution.

In principle, the sample could be biased due to the properties of their host galaxies.
We explore potential biases from host galaxies in Appendix~\ref{app:appA2}.
Physical properties of both PHANGS-ALMA and ACOS+archival galaxies share similar distributions of stellar mass, morphological type, and SFR; ALMA dataset properties are also compared (see Figs.~\ref{fig:figB1} and~\ref{fig:figB1_2}, and Tables~\ref{tab:tabA3} and~\ref{tab:tabA4} in Appendix~\ref{app:appA2sub1}).
Distances of the CCSN positions from the galactic centers are also analyzed in Appendix~\ref{app:appA2sub2} (see Figs.~\ref{fig:figB2} and~\ref{fig:figB3}), without a concentration within the central regions or outskirts.
Also, as a comparison sample the pixels at normalized distances lower than 0.3 (compared to the isophotal radius $D_{25}$, see Appendix~\ref{app:appA2}) are included.
Accordingly, when the comparison sample is limited to pixels at distances closer than $0.3 D_{25}$, none of the results change.

Residuals of the analyzed properties at the CCSN positions after subtracting the median values (with 1$\sigma$ confidence intervals) of the PHANGS-ALMA pixels at the same galactocentric radii were also calculated (Fig.~\ref{fig:figB2} and Table~\ref{tab:tabA5}).
The molecular surface density residual $\Delta \log \left( \Sigma_{\rm mol} / \rm{M}_{\odot}\,\rm{pc}^{-2} \right)$ is $0.51^{+0.12}_{-0.02}$ and $0.33^{+0.02}_{-0.02}$ dex for H-rich SNe and SESNe, respectively.
The velocity dispersion residual $\Delta  \log \left( \sigma_{\rm CO} / \rm{km}\,\rm{s}^{-1} \right)$ is $0.32^{+0.03}_{-0.05}$ and $0.34^{+0.01}_{-0.02}$ dex for H-rich SNe and SESNe, respectively.
The virial parameter residual $\Delta \log \left( \alpha_{\rm vir} \right)$ is $0.19^{+0.05}_{-0.07}$ and $0.53^{+0.09}_{-0.09}$ dex for H-rich and SESNe, respectively.
A similar analysis is repeated for the normalized distance from the galactic center (Fig.~\ref{fig:figB3} and Table~\ref{tab:tabA5}).
The computed values for H-rich SNe are $0.24^{+0.01}_{-0.03}$, $0.14^{+0.01}_{-0.01}$, and $0.27^{+0.07}_{-0.07}$ dex for $\Delta \log \left( \Sigma_{\rm mol} / \rm{M}_{\odot}\,\rm{pc}^{-2} \right)$, $\Delta  \log \left( \sigma_{\rm CO} / \rm{km}\,\rm{s}^{-1} \right)$, and $\Delta \log \left( \alpha_{\rm vir} \right)$, respectively.
For SESNe, this results in $0.10^{+0.03}_{-0.03}$, $0.28^{+0.02}_{-0.02}$, and $0.54^{+0.06}_{-0.06}$ dex for $\Delta \log \left( \Sigma_{\rm mol} / \rm{M}_{\odot}\,\rm{pc}^{-2} \right)$, $\Delta  \log \left( \sigma_{\rm CO} / \rm{km}\,\rm{s}^{-1} \right)$, and $\Delta \log \left( \alpha_{\rm vir} \right)$, respectively.

We also computed the excess of $\Sigma_{\rm mol}$--$\sigma_{\rm CO}$ ($\sigma_{\rm CO(2-1), \, Excess}$) and $\Sigma_{\rm mol}$--$\alpha_{\rm vir}$ ($\alpha_{\rm vir, \, Excess}$) comparing PHANGS-ALMA reference data and CCSN locations (Figure~\ref{fig:figB4} and Table~\ref{tab:tabA6}; Appendix~\ref{app:appA2sub3}).
The $\log \left( \sigma_{\rm CO(2-1), \, Excess} / \rm{km}\,\rm{s}^{-1} \right)$ is $0.11^{+0.01}_{-0.03}$ and $0.20^{+0.02}_{-0.01}$ dex for H-rich and SESNe, respectively.
$\log \left( \alpha_{\rm vir, \, Excess} \right)$ is $0.47^{+0.06}_{-0.10}$ and $0.61^{+0.08}_{-0.07}$ dex for H-rich and SESNe, respectively.
Furthermore, a similar analysis is performed for the galactic structures in the PHANGS-ALMA galaxies \citep{2021Querejeta} in Appendix~\ref{app:appA2sub4} (Figs.~\ref{fig:figB6_1}, ~ \ref{fig:figB6_2}, and~\ref{fig:figB6_3}), yielding a $\log \left( \sigma_{\rm CO} / \rm{km}\,\rm{s}^{-1} \right)$ median excess of $+$0.002 dex.

%%%%%%%%%%%%%%%%%%%%%%%%%%%%%%%%%%%%%%%%%%%%%%%%%%%%%%%%%%%%%%
\section{Discussion}\label{sec:sec4}

\subsection{High turbulence at the core-collapse supernova sites}

In this study, we investigate the molecular gas properties on molecular-cloud scales at CCSN explosion sites using a large sample of nearby galaxies.
The fact that CCSNe explode in high-density environments compared to other CO-detected positions of their hosts galaxies have been reported before \citep{2017Galbany,2023MaykerChen,2024Solar} and our findings are consistent with these previous results.
This effect is likely due to the higher probability of forming a SN progenitor in gas-rich regions, because they also have high star formation rates in accordance with the Kennicutt--Schmidt law \citep[or star-forming law;][]{1959Schmidt,1998Kennicutt}.

Previous studies of molecular gas in host galaxies of transients are not limited to SNe \citep{2017Galbany,2023MaykerChen,2024Solar}, but also concern fast radio bursts \citep{2018Bower,2022Hatsukade,2023Hsu,2023Chittidi,2024Yamanaka}, super-luminous SNe \citep{2019Arabsalmani,2020Hatsukade_b}, fast blue optical transients \citep{2019MorokumaMatsui,2023Sun}, broad-lined SNe Ic \citep{2018Michalowski_a,2020Michalowski}, and gamma-ray bursts \citep{2014Hatsukade,2016Michalowski,2018Arabsalmani,2018Michalowski_b,2020deUgartePostigo,2020Hatsukade_a,2021Chen,2023Nadolny,2024deUgartePostigo,2024Thone}.
Additionally, gas properties at the locations of RSG stars, Wolf-Rayet stars, and SN remnants were analyzed \citep{2023Sarbadhicary}, confirming our method to use molecular gas to constrain the nature of SN progenitors, as more massive (and short-lived) stars did exhibit higher molecular gas densities.
However, these works only focused on the integrated flux moment 0 maps (tracing the molecular gas mass), for a limited sample, and/or at kiloparsec resolution (tracing only the scales of spiral arms, disks, bulges, and bars).

We find an empirical correlation that CCSNe explode in highly turbulent molecular gas regions.
Given that the lifetimes of GMCs \citep{2022Kim} is comparable to the delay-time of CCSNe \citep{2017Zapartas,2026Souropanis}, these conditions reflect increased turbulence at the time when the progenitors were born.
High turbulence at the CCSN positions can be explained in three not mutually exclusive ways.

First, simply more turbulent regions tend to have more gas density (e.g.,~Fig.~\ref{fig:fig1}) at which the star formation, and hence, the probability of forming a massive star is higher.
Turbulent GMCs exhibit a wide range of properties, 
as gas is found at low densities, but also a significant quantity is at high densities \citep{2023Neumann}.
Turbulence as a key driver in regulating star formation and the formation of dense gas is also supported by \textit{JWST}\footnote{https://science.nasa.gov/mission/webb/} observations. 
They reveal that the edges of bubble structures are associated with young stellar populations, suggesting that stars continuously form in feedback-dominated systems, as gas is swept up and continues forming stars in the shells \citep{2023Barnes,2023aWatkins,2023bWatkins}.
Similarly, \citet{2026Fontaine} shows from numerical modeling of SN remnants that shocks affect the dense cloud, boosting star formation.
Indeed, previous generations of CCSNe could trigger a quasi-instantaneous formation of massive stars \citep{2016Padoan}.
% \textbf{One possibility of the dense molecular gas associated to the turbulent gas found at CCSN sites could be due to SN remnants enhancing the R$_{21}$ factor from shock–cloud interactions \citep{2013Jeong,2023Sano}.}
However, even at fixed gas surface density, the velocity dispersion at the explosion sites remains significantly higher than that at the reference regions (Fig.~\ref{fig:figB4}).
Therefore, this scenario alone cannot fully explain the observed trend.

Second, this may be connected with binary nature of CCSN progenitors.
More turbulence produces larger density fluctuations and these fluctuations enhance the binarity fraction, as fragmentation occurs intrinsically in these regions \citep{2004Goodwin,2007Goodwin, 2007McKee,2009Turk, 2015Konyves,2015Pineda,2023Kuruwita}.
In other words, the turbulent initial state of the gas leads to the formation of binary stars.
As a consequence, the vast majority of CCSNe exploding in molecular clouds dominated by non-thermal kinetic energy (i.e., clouds not in virial equilibrium nor free-fall) supports the idea that the progenitors of CCSNe form and evolve in massive binary stellar systems.
It is also proposed that that dynamical interaction (from large turbulence and density environments) plays an important role for the formation of massive binary stars \citep{1994Ostriker,2019Wall,2023Xu,2026Luo}.
If this interpretation is correct, then CCSNe can be used as tracers of the multiplicity fraction in turbulent environments at spatial scales of $\sim 100$\,pc, since turbulence at large scales is recovered at smaller ones due to a kinetic power law decay \citep{1941Kolmogorov,2020Traficante,2025Ma}.

Third, an alternative scenario is that CCSNe in our sample did not explode as the first generation of stars formed from the same molecular cloud and the turbulence could be due to prior CCSNe having formed at the same time but exploded earlier.
If this is the case, then high turbulence is not representative of the state of the GMC at the birth of the progenitor star and is not connected with the possibility of the high density and/or binarity fraction.
Around 20\% of the binary massive stars are expected to move from their birth sites due to a disruption by a supernova kick \citep{2019Renzo}.
These ejected companions, the mean distance traveled is $\sim$\,$100\,$pc, without a connection with the GMC parent in study.
This means that a continuous source of converging flows maintains turbulence within galaxies and this could be sustained by previous generations of CCSNe \citep{2020Bacchini,2020Lu}, consistent with a  ``turbulent support" scenario for molecular clouds and star formation \citep{2026VazquezSemadeni}.

This increases the turbulence at the time of explosion only for the lower-mass CCSN progenitors as their delay-times are longer. The environment of higher-mass progenitors at the time of explosion reflects the condition at birth, because they explode first.
It is not easy to test this effect, as the connection of the current turbulence and binarity at birth is open to discussion.

On the other hand, the studied CCSNe do not contribute to the turbulence detected within the molecular gas at GMC scales.
Assuming a constant expanding shock velocity of 10\,000\,km\,s$^{-1}$ as an upper limit and using the oldest CCSN from our sample (having exploded $\sim 70$ years ago), the shock would reach a diameter of merely 1.5\,pc \citep{2011Draine}.
This means that the shock-cloud interaction influences at most $\sim 3.0 \times 10^{-6}$ of the turbulence observed in the volume of the GMCs, being a negligible contribution.
Higher resolutions would be required in order to understand the input to the turbulence of these CCSNe.
Similarly, the pre-SN feedback of the progenitors of SNe studied here could not influence our result.
A star can ionize gas around it up to the so-called Str{\"o}mgren radius, which can be calculated as $(3Q_H/4\pi\alpha_B n^2)^{1/3}$ \citep{Stromgren39}, where $Q_H=10^{48-50}\,\mbox{s}^{-1}$ is the ionizing photon rate for O-type stars \citep{Panagia73,Vacca96,Martins05}, likely progenitors of CCSNe; $\alpha_B=2.59\times10^{-13}\,\mbox{cm}^3\,\mbox{s}^{-1}$ is the case B recombination coefficient for the temperature of 10\,000\,K \citep{Osterbrock89}; and $n$ is the hydrogen number density assumed to be $100\,\mbox{cm}^{-3}$ for a typical molecular cloud. This results in a Str{\"o}mgren radius of $1.5$--$6.8$\,pc, again too small to affect our measurements at the 100\,pc scale.
Finally, the mechanical pre-SN feedback affects at most 30 pc \citep[][fig.~5]{Fichtner2024}, which corresponds only to $(30/100)^3=2.7\%$ of the volume probed by our observations.

While studies from pre-explosion images --- complemented with follow-up observations confirming the disappearance of CCSN progenitors --- find companions for a few SESNe \citep{2026Zapartas}, the lack of detected companions for H-rich CCSN progenitors still allows a binary nature.
This is the case if the companion is too faint or for binary systems which eventually evolve into isolated progenitors after merging or after ejection of the companion prior to the SN explosion \citep{2019Zapartas}.
% Another case is \object{SN2023ixf}, one of the most nearby SNe ($\sim 7$\,Mpc away), which exploded in a very dusty environment \citep{2023Kilpatrick,2024VanDyk,2024Xiang}, so that \cite{2024Qin} showed the possibility of a companion of mass below 5.6\,M$_{\sun}$.
Another case is \object{SN2023ixf} (one of the most nearby extragalactic SN detected, $\sim\,$$7$\,Mpc away), for which \cite{2024Qin} constrained an upper limit on the companion's mass of 5.6\,M$_{\sun}$, accounting for a high dust extinction \citep{2023Kilpatrick,2024VanDyk,2024Xiang}.
Moreover, Betelgeuse, a red supergiant star (RSG; similar to progenitors of H-rich SNe) located $\sim 168$\,pc away could have a lower-mass stellar companion based on direct-imaging detection \citep{2025Goldberg,2025Howell}.
Similarly, \object{SN1987A}’s \citep[][and references therein]{1989Arnett} progenitor is thought to have been a blue supergiant star formed in a stellar merger, supported by observational and theoretical studies \citep{1992bPodsiadlowski,2007Morris}.
{In summary, the non-detection of companions of SN progenitors may be due to historical binary interaction \citep[e.g., merger, unbounded system;][]{2026Zapartas} or instrumental limitations.

\subsection{Caveats}

The sample analyzed here includes nearby CCSNe ($< 100$\,Mpc; due to instrumental limitations) and the host galaxies are normal, star forming galaxies found on the main sequence of star formation. 
Hence, our conclusions concern the Local Universe, and the conditions of star formation at earlier epochs may be different.

There is no consensus how active galactic nuclei (AGN) activity significantly influences the cold gas content in their host galaxies \citep{1998Boyle,2011Sturm,2013Merloni,2014Cicone,2026Chen}.
A total of seven (out of 27) contain an AGN \citep[classified by optical spectroscopy;][]{2014Makarov}.
We note that there is an overlap of the parameters in this study (normalized distance and distance from galactic center, $\Sigma_{\rm mol}$, $\sigma_{\rm CO}$, and $\alpha_{\rm vir}$) between host galaxies with and without AGNs (see Figs.~\ref{fig:figB2} and~\ref{fig:figB3}).
Although AGN feedback is expected to enhance the local turbulence, both $\alpha_{\rm{CO}}^{1-0}$ and R$_{21}$ remain similar for host galaxies of low-redshift quasars and normal star-forming galaxies \citep{2020Shangguan, 2020bShangguan}.
If AGN hosts are removed from the sample, the trends remain unchanged.

Another caveat is that we used only strong detections of CO(2--1) at $S/N > 4$.
This means that the sample may be biased towards bright and turbulent molecular gas regions, excluding the faint and less turbulent ones, as expected from the Larson's law \citep{1981Larson}.
However, fluxes lower than those at the positions of most of the CCSNe are also detected throughout our host galaxies.
On the other hand, the high signal-to-noise threshold used in the observations give us a low rate of false positive signals from CO(2--1), that could be contaminated by noise.

A further point to consider is that the data are taken from different datasets (PHANGS, ACOS, and archival observations), which have different beam sizes and different levels of missing flux.
We note that around 50\% of the total CO flux could be missing using a single ALMA configuration of the antennas (12 m or 7 m data) instead of 12m+7m+TP data \citep{2013Pety}.
As 11 (1) out of the 33 CCSNe are observed with 12 m-only (7 m-only) data, their fluxes may be underestimated compared to those of the reference regions measured from the PHANGS 12m+7m+TP data.
If this is the case, our conclusion that CCSN sites have higher $\Sigma_{\rm mol}$ and $\sigma_{\rm CO}$ than the reference regions would be even stronger.
% As 11 (1) out of 33 CCSNe include single 12 m (7 m) data, fluxes could be underestimated compared to host galaxy pixels, which makes our conclusions even stronger.
% In parallel, it is importance to notice that the sample PHANGS-ALMA has higher sensitivities (but similar beam sizes) than ACOS+archival data (Fig.~\ref{fig:figB1_2}), hence, there is an observational bias shifter toward lower values of pixels for the former.}
In parallel, it is important to notice that the data of PHANGS-ALMA and ACOS+archival are statistically consistent in terms of sensitivities (Fig.~\ref{fig:figB1_2}).
The p-values of 0.059 and 0.012 corresponds (Table~\ref{tab:tabA4}) to 1.9$\sigma$ and 2.5$\sigma$ difference, respectively.
Hence, we rule out an observational bias for the selection effect.
We used data at the best resolution available and an ideal sample should contain similar values of spatial and velocity resolutions.
However, if the spatial resolution is decreased to the coarsest one used here (i.e., 229 pc), molecular gas might be contaminated with structures not associated with the parent molecular gas at GMC scales of the CCSN.
A balance is to make use of native resolutions from ALMA in order to extract the best properties at cloud-scale observations.
Finally, the ideal tracers of cold dense gas are HCN(1--0), HCO$^{+}$(1--0), and CS(2--1) lines \citep{2023Neumann}, but they are fainter, so are usually detected only with lower resolutions (typically kiloparsecs).

%%%%%%%%%%%%%%%%%%%%%%%%%%%%%%%%%%%%%%%%%%%%%%%%%%%%%%%%%%%%%%
\section{Conclusions}\label{sec:sec5}

In this Paper, we have analyzed the molecular gas properties, $\Sigma_{\rm{mol}}$, and, for the first time, $\sigma_{\rm{CO}}$ and $\alpha_{\rm{vir}}$, for 19 H-rich SNe and 14 SESNe (total of 33 CCSNe).
We used only pixels with strong detections %of signal-to-noise 
($S/N > 4$) at spatial scales of $\sim 100$\,pc, with the aim of constraining the nature of the CCSN progenitors.
We summarize our results as follows:

\begin{itemize}
    \item Both CCSN populations explode in similar conditions of the molecular gas surface density, but for higher $\Sigma_{\rm{mol}}$ values in comparison with the average of their host galaxies,
    \item CCSNe are located in molecular gas regions with higher turbulence than the average of their host galaxies, and
    \item SESNe are found in slightly more turbulent molecular gas than H-rich SNe (evidenced by the dynamical state of GMCs using the virial parameter), and more data is required to obtain statistical significance.
\end{itemize}

The fact that CCSNe explode in turbulent molecular gas regions, gives us insights into their progenitors and three different scenarios are proposed.
1) Turbulence enhances gas density and dense-gas formation in molecular gas at GMC scales, thereby promoting star formation and the production of massive stars, a picture supported by both observations and simulations showing that feedback-driven shocks and shells can trigger successive generations of star formation.
2) As the stellar multiplicity fraction is expected to increase in dense regions (due to turbulent motions), this could suggest that CCSN progenitors are dominated by binary systems.
Furthermore, the formation of interacting systems (SESN progenitors) might be enhanced in turbulent regions, in comparison with wide binaries (H-rich SN progenitors).
3) An alternative interpretation is that CCSNe born from the same gas as those from our sample but having exploded earlier could increase the turbulence in the regions for the CCSNe with the longest delay-times (lowest initial masses).

%%%%%%%%%%%%%%%%%%%%%%%%%%%%%%%%%%%%%%%%%%%%%%%%%%%%%%%%%%%%%%
\begin{acknowledgements}
    We thank Cecilia Bacchini for valuable comments.
    This research was funded in whole or in part by the National Science Centre (NCN), Poland, projects PRELUDIUM 2024/53/N/ST9/00350 and OPUS 2023/49/B/ST9/00066.
    For the purpose of Open Access, the author has applied a CC-BY public copyright licence to any Author Accepted Manuscript (AAM) version arising from this submission.
    Supported by the Foundation for Polish Science (FNP). J.N.~acknowledges the support of the Polish National Agency for Academic Exchange (NAWA) Bekker grant BPN/BEK/2023/1/00271, and the kind hospitality of the IAC.
    L.G. acknowledges financial support from AGAUR, CSIC, MCIN and AEI 10.13039/501100011033 under projects PID2023-151307NB-I00, PIE 20215AT016, CEX2020-001058-M, ILINK23001, COOPB2304, and 2021-SGR-01270.
    This work was supported by a research grant (VIL54489) from VILLUM FONDEN.
    This research has made use of the Transient Name Server (TNS), operated by the IAU Supernova Working Group and hosted by the Weizmann Institute of Science.
    We used IAU Circulars and CBETs presented by the Central Bureau for Astronomical Telegrams.
    We acknowledge the use of the Supernova Catalog maintained by the Institute of Astronomy, Moscow State University (SAI MSU).
    The catalog can be accessed at: http://stella.sai.msu.ru/sncat/.
    We acknowledge the work of astronomers supporting the LEDA database (http://leda.univ-lyon1.fr/), the Asiago Supernova Catalogue (http://graspa.oapd.inaf.it/asnc.html), The STScI Digitized Sky Survey (http://archive.stsci.edu/cgi-bin/dss\_form), and CfA List of Supernovae (http://www.cbat.eps.harvard.edu/lists/RecentSupernovae.html).
    This research has made use of the NASA/IPAC Extragalactic Database (NED) which is operated by the Jet Propulsion Laboratory, California Institute of Technology, under contract with the National Aeronautics and Space Administration (http://ned.ipac.caltech.edu/).
    ALMA is a partnership of ESO (representing its member states), NSF (USA) and NINS (Japan), together with NRC (Canada), MOST and ASIAA (Taiwan), and KASI (Republic of Korea), in cooperation with the Republic of Chile.
    The Joint ALMA Observatory is operated by ESO, AUI/NRAO and NAOJ.
    The National Radio Astronomy Observatory is a facility of the National Science Foundation operated under cooperative agreement by Associated Universities, Inc.
    This paper makes use of the following ALMA data:
    ADS/JAO.ALMA\#2013.1.01161.S : PI Sakamoto (archival), NGC\,1365, NGC\,5236 (M83); 
    ADS/JAO.ALMA\#2015.1.00121.S : PI Sakamoto (archival), NGC\,5236 (M83);
    ADS/JAO.ALMA\#2015.1.00902.S : PI Ao, PGC\,084885;
    ADS/JAO.ALMA\#2015.1.00925.S : PI Blanc (pilot), NGC\,1087, NGC\,1566; 
    ADS/JAO.ALMA\#2015.1.00956.S : PI Leroy (pilot), NGC\,1672, NGC\,3627, NGC\,4254, NGC\,4303, NGC\,4321, NGC\,6744;
    ADS/JAO.ALMA\#2016.1.00386.S : PI Sakamoto (archival), NGC\,5236 (M83);
    ADS/JAO.ALMA\#2017.1.00255.S : PI Pereira Santaella , NGC\,1614;
    ADS/JAO.ALMA\#2017.1.00392.S : PI Blanc (pilot follow up), NGC\,1087, NGC\,1566;
    ADS/JAO.ALMA\#2017.1.00886.L : PI Schinnerer (large program), NGC\,1097, NGC\,1559, NGC\,1637, NGC\,2997, NGC\,5530;
    ADS/JAO.ALMA\#2018.A.00062.S : PI Faesi (very nearby extension), NGC\,5128; 
    ADS/JAO.ALMA\#2018.1.00272.S : PI Sliwa, NGC\,4038;
    ADS/JAO.ALMA\#2018.1.00538.S : PI Izumi, NGC\,5231;
    ADS/JAO.ALMA\#2018.1.01651.S : PI Leroy (pilot follow up), NGC\,1087, NGC\,1566;
    ADS/JAO.ALMA\#2019.1.01305.S : PI Tanaka, NGC\,4981;
    ADS/JAO.ALMA\#2021.1.00099.S : PI Michałowski (ACOS survey), ESO\,492-002, NGC\,0908, NGC\,3278, NGC\,3354, NGC\,4219, NGC\,5967.
\end{acknowledgements}

%%%%%%%%%%%%%%%%%%%%%%%%%%%%%%%%%%%%%%%%%%%%%%%%%%%%%%%%%%%%%%
\bibliographystyle{aa} % style aa.bst
\bibliography{biblio} % your references Yourfile.bib

%%%%%%%%%%%%%%%%%%%%%%%%%%%%%%%%%%%%%%%%%%%%%%%%%%%%%%%%%%%%%%
\begin{appendix}
\onecolumn

%%%%%%%%%%%%%%%%%%%%%%%%%%%%%%%%%%%%%%%%%%%%%%%%%%%%%%%%%%%%%%
\section{CCSN properties and environments}\label{app:appA1}

Table~\ref{tab:tabA1} presents the CCSN properties, extracted from different databases.
Table~\ref{tab:tabA2} shows the CCSN environments characterized in this study.

\begin{table*}[ht!]
    \caption{CCSN properties}
    \label{tab:tabA1}
    \centering
    \begin{tabular}{ccccccc}
        \hline
        \hline
        Name & R.A. & Decl. & Type & Group & Redshift & Host galaxy name\\
        {} & hms & \degr,\arcmin,\arcsec & {} & {} & {} & {}\\
        \hline
        \object{ASASSN-14ha} & 4$^{\rm{h}}$20$^{\rm{m}}$01$^{\rm{s}}$.41 & -54\degr56\arcmin17\arcsec.00 & II & H-rich SNe & 0.005017 & \object{NGC1566} \\
        \object{SN1964F} & 12$^{\rm{h}}$21$^{\rm{m}}$53$^{\rm{s}}$.02 & 04\degr28\arcmin24\arcsec.30 & II & H-rich SNe & 0.005225     & \object{NGC4303} \\
        \object{SN1967H} & 12$^{\rm{h}}$18$^{\rm{m}}$55$^{\rm{s}}$.12 & 14\degr24\arcmin40\arcsec.30 & II & H-rich SNe & 0.008023     & \object{NGC4254} \\
        \object{SN1968L} & 13$^{\rm{h}}$37$^{\rm{m}}$00$^{\rm{s}}$.51 & -29\degr51\arcmin59\arcsec.00 & IIP & H-rich SNe & 0.00169    & \object{NGC5236} \\
        \object{SN1972Q} & 12$^{\rm{h}}$18$^{\rm{m}}$50$^{\rm{s}}$.64 & 14\degr26\arcmin36\arcsec.30 & IIP & H-rich SNe & 0.008023    & \object{NGC4254} \\
        \object{SN1973R} & 11$^{\rm{h}}$20$^{\rm{m}}$11$^{\rm{s}}$.65 & 12\degr59\arcmin55\arcsec.00 & IIP & H-rich SNe & 0.002404    & \object{NGC3627} \\
        \object{SN1983V} & 3$^{\rm{h}}$33$^{\rm{m}}$31$^{\rm{s}}$.63 & -36\degr08\arcmin55\arcsec.00 & Ic & SESNe & 0.005462           & \object{NGC1365} \\
        \object{SN1986I} & 12$^{\rm{h}}$18$^{\rm{m}}$52$^{\rm{s}}$.04 & 14\degr24\arcmin44\arcsec.10 & IIP & H-rich SNe & 0.008023    & \object{NGC4254} \\
        \object{SN1992bd} & 2$^{\rm{h}}$46$^{\rm{m}}$19$^{\rm{s}}$.23 & -30\degr16\arcmin39\arcsec.10 & II & H-rich SNe & 0.004231    & \object{NGC1097} \\
        \object{SN1994ai} & 2$^{\rm{h}}$23$^{\rm{m}}$06$^{\rm{s}}$.17 & -21\degr13\arcmin58\arcsec.30 & Ic & SESNe & 0.005026          & \object{NGC0908} \\
        \object{SN1995V} & 2$^{\rm{h}}$46$^{\rm{m}}$26$^{\rm{s}}$.77 & -00\degr29\arcmin55\arcsec.60 & II & H-rich SNe & 0.005074     & \object{NGC1087} \\        
        \object{SN1999em} & 4$^{\rm{h}}$41$^{\rm{m}}$27$^{\rm{s}}$.13 & -02\degr51\arcmin45\arcsec.40 & IIP & H-rich SNe & 0.002388   & \object{NGC1637} \\
        \object{SN1999gn} & 12$^{\rm{h}}$21$^{\rm{m}}$57$^{\rm{s}}$.02 & 04\degr27\arcmin45\arcsec.60 & IIP & H-rich SNe & 0.005225   & \object{NGC4303} \\
        \object{SN2003jg} & 9$^{\rm{h}}$45$^{\rm{m}}$37$^{\rm{s}}$.91 & -31\degr11\arcmin21\arcsec.00 & Ib/c & SESNe & 0.003626        & \object{NGC2997} \\
        \object{SN2004ip} & 18$^{\rm{h}}$32$^{\rm{m}}$41$^{\rm{s}}$.26 & -34\degr11\arcmin26\arcsec.70 & II & H-rich SNe & 0.018276   & \object{PGC084885} \\
        \object{SN2005at} & 19$^{\rm{h}}$09$^{\rm{m}}$53$^{\rm{s}}$.57 & -63\degr49\arcmin22\arcsec.80 & Ic & SESNe & 0.002803         & \object{NGC6744} \\
        \object{SN2005lr} & 7$^{\rm{h}}$11$^{\rm{m}}$39$^{\rm{s}}$.03 & -26\degr42\arcmin20\arcsec.20 & Ic & SESNe & 0.008647          & \object{ESO492-002} \\
        \object{SN2007C} & 13$^{\rm{h}}$08$^{\rm{m}}$49$^{\rm{s}}$.30 & -06\degr47\arcmin01\arcsec.00 & Ib & SESNe & 0.005595          & \object{NGC4981} \\
        \object{SN2007it} & 14$^{\rm{h}}$18$^{\rm{m}}$25$^{\rm{s}}$.57 & -43\degr22\arcmin54\arcsec.20 & II & H-rich SNe & 0.003966   & \object{NGC5530} \\
        \object{SN2009bb} & 10$^{\rm{h}}$31$^{\rm{m}}$33$^{\rm{s}}$.87 & -39\degr57\arcmin30\arcsec.00 & Ic-BL & SESNe & 0.009995     & \object{NGC3278} \\
        \object{SN2009gd} & 15$^{\rm{h}}$48$^{\rm{m}}$22$^{\rm{s}}$.79 & -75\degr40\arcmin47\arcsec.80 & Ic & SESNe & 0.009745         & \object{NGC5967} \\
        \object{SN2009hd} & 11$^{\rm{h}}$20$^{\rm{m}}$16$^{\rm{s}}$.96 & 12\degr58\arcmin46\arcsec.60 & II & H-rich SNe & 0.002404    & \object{NGC3627} \\
        \object{SN2009ib} & 4$^{\rm{h}}$17$^{\rm{m}}$39$^{\rm{s}}$.92 & -62\degr46\arcmin38\arcsec.70 & IIP & H-rich SNe & 0.004319   & \object{NGC1559} \\
        \object{SN2011hp} & 12$^{\rm{h}}$16$^{\rm{m}}$25$^{\rm{s}}$.47 & -43\degr19\arcmin46\arcsec.90 & Ic & SESNe & 0.00662          & \object{NGC4219} \\
        \object{SN2011jl} & 10$^{\rm{h}}$43$^{\rm{m}}$02$^{\rm{s}}$.95 & -36\degr21\arcmin52\arcsec.40 & Ic & SESNe & 0.010001         & \object{NGC3354} \\
        \object{SN2013dk} & 12$^{\rm{h}}$01$^{\rm{m}}$52$^{\rm{s}}$.72 & -18\degr52\arcmin18\arcsec.30 & Ic & SESNe & 0.00548          & \object{NGC4038} \\
        \object{SN2014L} & 12$^{\rm{h}}$18$^{\rm{m}}$48$^{\rm{s}}$.71 & 14\degr24\arcmin44\arcsec.40 & Ic & SESNe & 0.008023           & \object{NGC4254} \\
        \object{SN2016adj} & 13$^{\rm{h}}$25$^{\rm{m}}$24$^{\rm{s}}$.11 & -43\degr00\arcmin57\arcsec.90 & IIb & SESNe & 0.001889       & \object{NGC5128} \\
        \object{SN2016cok} & 11$^{\rm{h}}$20$^{\rm{m}}$19$^{\rm{s}}$.10 & 12\degr58\arcmin56\arcsec.00 & IIP & H-rich SNe & 0.002404  & \object{NGC3627} \\
        \object{SN2020cuj} & 4$^{\rm{h}}$34$^{\rm{m}}$00$^{\rm{s}}$.53 & -08\degr34\arcmin43\arcsec.46 & II & H-rich SNe & 0.0158     & \object{NGC1614} \\
        \object{SN2020oi} & 12$^{\rm{h}}$22$^{\rm{m}}$54$^{\rm{s}}$.93 & 15\degr49\arcmin24\arcsec.96 & Ic & SESNe & 0.0052            & \object{NGC4321} \\        
        \object{SN2021kwc} & 13$^{\rm{h}}$35$^{\rm{m}}$47$^{\rm{s}}$.918 & 02\degr59\arcmin59\arcsec.35 & IIn & H-rich SNe & 0.021759 & \object{NGC5231} \\        
        \object{SN2022aau} & 4$^{\rm{h}}$45$^{\rm{m}}$41$^{\rm{s}}$.75 & -59\degr14\arcmin43\arcsec.38 & II & H-rich SNe & 0.00444    & \object{NGC1672} \\
        \hline
    \end{tabular}
\end{table*}

\begin{table*}[ht!]
    \caption{CCSN location values computed in this work.}
    \label{tab:tabA2}
    \centering
    \begin{tabular}{ccccccccc}
        \hline
        \hline
        Name & Group & $\Sigma_{\rm{mol}}$ & $\sigma_{\rm{CO}(2-1)}$ & $\alpha_{\rm{vir}}$ & r$_{\rm{SN}}$ & r$_{\rm{SN,norm}}$\\
        {} & {} & (M$_{\sun}$ pc$^{-2}$) & (km s$^{-1}$) & {} & (kpc) &{} \\
        \hline
        \object{ASASSN-14ha} & H-rich SNe & 256.8 $\pm$ 3.0    & 11.7 $\pm$ 0.1  & 3.7 $\pm$ 0.1     & 0.9  & 0.04 \\
        \object{SN1964F}     & H-rich SNe & 72.6 $\pm$ 2.9     & 20.7 $\pm$ 0.5  & 29.2 $\pm$ 1.8    & 3.0  & 0.13 \\
        \object{SN1967H}     & H-rich SNe & 140.4 $\pm$ 2.0    & 5.4 $\pm$ 0.1   & 1.4 $\pm$ 0.1     & 13.9 & 0.53 \\
        \object{SN1968L}     & H-rich SNe & 688.5 $\pm$ 5.1    & 28.7 $\pm$ 4.0  & 17.5 $\pm$ 4.8    & 0.2  & 0.01 \\
        \object{SN1972Q}     & H-rich SNe & 4.2 $\pm$ 1.2      & 1.3 $\pm$ 0.3   & 2.5 $\pm$ 1.2     & 16.4 & 0.62 \\
        \object{SN1973R}     & H-rich SNe & 41.8 $\pm$ 1.5     & 9.6 $\pm$ 2.4   & 18.4 $\pm$ 9.0    & 2.8  & 0.18 \\
        \object{SN1983V}     & SESNe      & 9.3 $\pm$ 1.3      & 3.4 $\pm$ 1.3   & 7.0 $\pm$ 5.3     & 7.4  & 0.18 \\
        \object{SN1986I}     & H-rich SNe & 28.1 $\pm$ 2.0     & 8.0 $\pm$ 0.3   & 14.8 $\pm$ 1.5    & 6.6  & 0.25 \\
        \object{SN1992bd}    & H-rich SNe & 450.1 $\pm$ 1.9    & 19.7 $\pm$ 0.1  & 5.7 $\pm$ 0.0     & 0.8  & 0.03 \\
        \object{SN1994ai}    & SESNe      & 6.0 $\pm$ 2.5      & 10.4 $\pm$ 1.5  & 269.8 $\pm$ 134.6 & 2.4  & 0.13 \\
        \object{SN1995V}     & H-rich SNe & 38.7 $\pm$ 1.1     & 3.8 $\pm$ 0.1   & 2.2 $\pm$ 0.1     & 2.6  & 0.27 \\
        \object{SN1999em}    & H-rich SNe & 10.4 $\pm$ 0.8     & 3.8 $\pm$ 1.1   & 13.2 $\pm$ 8.0    & 1.2  & 0.24 \\
        \object{SN1999gn}    & H-rich SNe & 30.1 $\pm$ 2.6     & 8.6 $\pm$ 0.3   & 12.1 $\pm$ 1.3    & 5.7  & 0.25 \\
        \object{SN2003jg}    & SESNe      & 156.9 $\pm$ 1.5    & 13.8 $\pm$ 0.1  & 7.5 $\pm$ 0.1     & 1.0  & 0.04 \\
        \object{SN2004ip}    & H-rich SNe & 1772.6 $\pm$ 160.1 & 49.7 $\pm$ 0.7  & 10.5 $\pm$ 1.0    & 0.7  & 0.08 \\
        \object{SN2005at}    & SESNe      & 16.5 $\pm$ 1.7     & 2.5 $\pm$ 0.8   & 5.6 $\pm$ 3.7     & 7.9  & 0.28 \\
        \object{SN2005lr}    & SESNe      & 76.9 $\pm$ 7.4     & 9.7 $\pm$ 0.2   & 9.6 $\pm$ 1.0     & 3.3  & 0.23 \\
        \object{SN2007C}     & SESNe      & 32.3 $\pm$ 1.9     & 8.9 $\pm$ 0.1   & 11.9 $\pm$ 0.8    & 2.8  & 0.28 \\
        \object{SN2007it}    & H-rich SNe & 2.5 $\pm$ 0.7      & 1.3 $\pm$ 0.1   & 7.2 $\pm$ 2.4     & 2.8  & 0.21 \\
        \object{SN2009bb}    & SESNe      & 167.9 $\pm$ 10.5   & 10.1 $\pm$ 0.2  & 4.9 $\pm$ 0.3     & 4.6  & 0.5 \\
        \object{SN2009gd}    & SESNe      & 15.7 $\pm$ 4.4     & 9.1 $\pm$ 0.8   & 37.0 $\pm$ 12.1   & 7.5  & 0.47 \\
        \object{SN2009hd}    & H-rich SNe & 355.8 $\pm$ 1.7    & 21.5 $\pm$ 0.1  & 10.8 $\pm$ 0.1    & 2.6  & 0.17 \\
        \object{SN2009ib}    & H-rich SNe & 2.3 $\pm$ 0.6      & 1.3 $\pm$ 0.1   & 4.4 $\pm$ 1.5     & 3.3  & 0.29 \\
        \object{SN2011hp}    & SESNe      & 18.7 $\pm$ 1.8     & 7.1 $\pm$ 0.2   & 18.1 $\pm$ 1.9    & 3.9  & 0.23 \\
        \object{SN2011jl}    & SESNe      & 9.3 $\pm$ 2.9      & 8.7 $\pm$ 0.6   & 56.4 $\pm$ 19.4   & 1.6  & 0.25 \\
        \object{SN2013dk}    & SESNe      & 60.7 $\pm$ 29.6    & 8.7 $\pm$ 1.5   & 21.1 $\pm$ 12.5   & 1.5  & 0.08 \\
        \object{SN2014L}     & SESNe      & 322.4 $\pm$ 2.1    & 91.4 $\pm$ 0.5  & 169.5 $\pm$ 2.3   & 3.5  & 0.13 \\
        \object{SN2016adj}   & SESNe      & 60.5 $\pm$ 3.2     & 18.3 $\pm$ 3.5  & 17.9 $\pm$ 6.9    & 1.8  & 0.05 \\
        \object{SN2016cok}   & H-rich SNe & 2.2 $\pm$ 0.6      & 1.3 $\pm$ 0.3   & 6.0 $\pm$ 3.6     & 3.5  & 0.22 \\
        \object{SN2020cuj}   & H-rich SNe & 400.3 $\pm$ 170.7  & 119.7 $\pm$ 5.0 & 479.5 $\pm$ 208.3 & 2.5  & 0.18 \\
        \object{SN2020oi}    & SESNe      & 105.3 $\pm$ 2.6    & 28.7 $\pm$ 0.5  & 47.1 $\pm$ 2.0    & 0.5  & 0.03 \\
        \object{SN2021kwc}   & H-rich SNe & 36.3 $\pm$ 15.0    & 11.9 $\pm$ 3.6  & 26.5 $\pm$ 19.3   & 2.6  & 0.16 \\
        \object{SN2022aau}   & H-rich SNe & 351.8 $\pm$ 4.2    & 20.4 $\pm$ 0.2  & 4.8 $\pm$ 0.1     & 0.8  & 0.05 \\
        \hline
    \end{tabular}
\end{table*}

%%%%%%%%%%%%%%%%%%%%%%%%%%%%%%%%%%%%%%%%%%%%%%%%%%%%%%%%%%%%%%
\section{Host galaxy dependence}\label{app:appA2}

\subsection{Dataset}\label{app:appA2sub1}

% In order to check if the host galaxy characteristics (such as stellar mass or morphology) are different for PHANGS-ALMA and ACOS+archival galaxies, we compare their physical properties.
% To do so, we used their apparent total K magnitudes (converted to absolute magnitudes, $M_{\rm{K}}$) and morphological type codes from HyperLEDA\footnote{http://atlas.obs-hp.fr/hyperleda/} \citep{2014Makarov} as estimators of stellar mass \citep{2003Bell} and galaxy morphology \citep{1991deVaucouleurs}, respectively.
% Figure~\ref{fig:figB1} shows the histograms of the stellar masses and morphological type codes comparing the PHANGS-ALMA and ACOS+archival galaxies.
% In addition, in Table~\ref{tab:tabA3} we present their values \textbf{\citep[and other parameters such as luminosity distance $D_{L}$, $M_{*}$, SFR, $i$, number of CCSNe, and corresponding ALMA ID;][]{2019Leroy}}.
% We also computed the KS and AD two--sample tests comparing their distributions (see Table~\ref{tab:tabA4}).\

In order to check if the host galaxy characteristics are different for PHANGS-ALMA and ACOS+archival galaxies, we compare their physical properties.
To do so, we used their apparent total K magnitudes (converted to absolute magnitudes, $M_{\rm{K}}$) and morphological type codes from HyperLEDA\footnote{http://atlas.obs-hp.fr/hyperleda/} \citep{2014Makarov} as estimators of stellar mass \citep{2003Bell} and galaxy morphology \citep{1991deVaucouleurs}, respectively.
Figure~\ref{fig:figB1} shows the histograms of $M_{\rm{K}}$, morphological type codes, SFR, and M$_{*}$ comparing the PHANGS-ALMA and ACOS+archival galaxies.
A similar analysis is performed for the ALMA datasets, shown in Fig.~\ref{fig:figB1_2} for the spatial resolution ($\theta$), velocity resolution ($\Delta v_{\rm{chan}}$), and 1$\sigma$ mass surface density sensitivity  ($\Sigma_{\rm{mol, 1\sigma}}$).
In addition, in Table~\ref{tab:tabA3} we present their values \citep[and other parameters such as luminosity distance $D_{L}$, $i$, $\theta$, number of CCSNe, and corresponding ALMA ID;][]{2019Leroy}.
% \textbf{As data from PHANGS-ALMA and ACOS+archival were constrained with different goals, ALMA observing parameters () are also shown in Fig.~\ref{fig:figB1_2}}
We also computed the KS and AD two-sample tests comparing their distributions (see Table~\ref{tab:tabA4}).

\begin{figure*}[ht!]
    \centering
    \resizebox{\hsize}{!}{\includegraphics{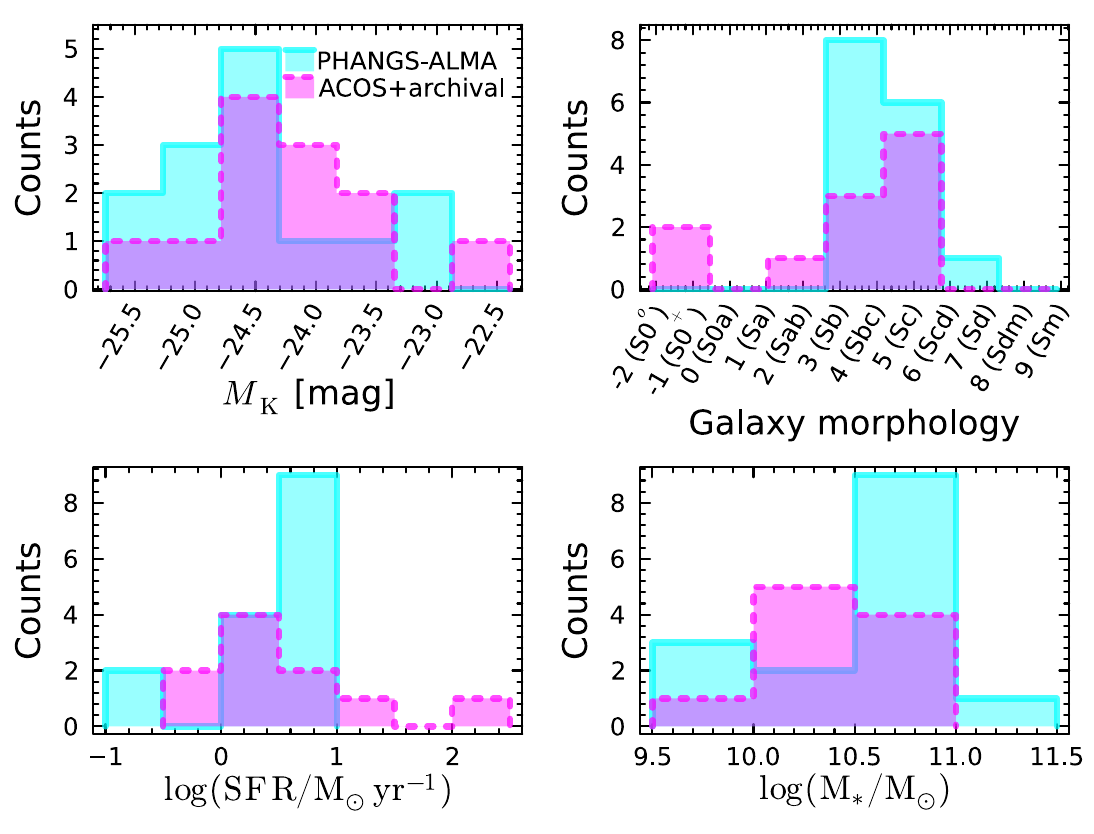}}
    \caption{Histogram comparing host galaxies from the PHANGS-ALMA sample in solid cyan and ACOS+archival data in dotted magenta.
    The values are the K-band absolute magnitude (top left panel), morphological type code (top right panel), star formation rate (bottom left panel), and stellar mass (bottom right panel)}.
    \label{fig:figB1}
\end{figure*}

From the KS (AD) tests we obtained p-values of 0.86 (0.72), 0.73 (0.48), 0.90 (0.64), and 0.65 (0.79) for $M_{\rm{K}}$, galaxy morphology, SFR, and M$_{*}$, respectively. Hence,
PHANGS-ALMA and ACOS+archival galaxies are in agreement, and 
it is not possible to reject the null hypothesis that both samples come from the same distributions.
Hence, the physical properties of the host galaxies in this study do not bias the CCSN sample.

\begin{figure*}[ht!]
    \centering
    \resizebox{\hsize}{!}{\includegraphics{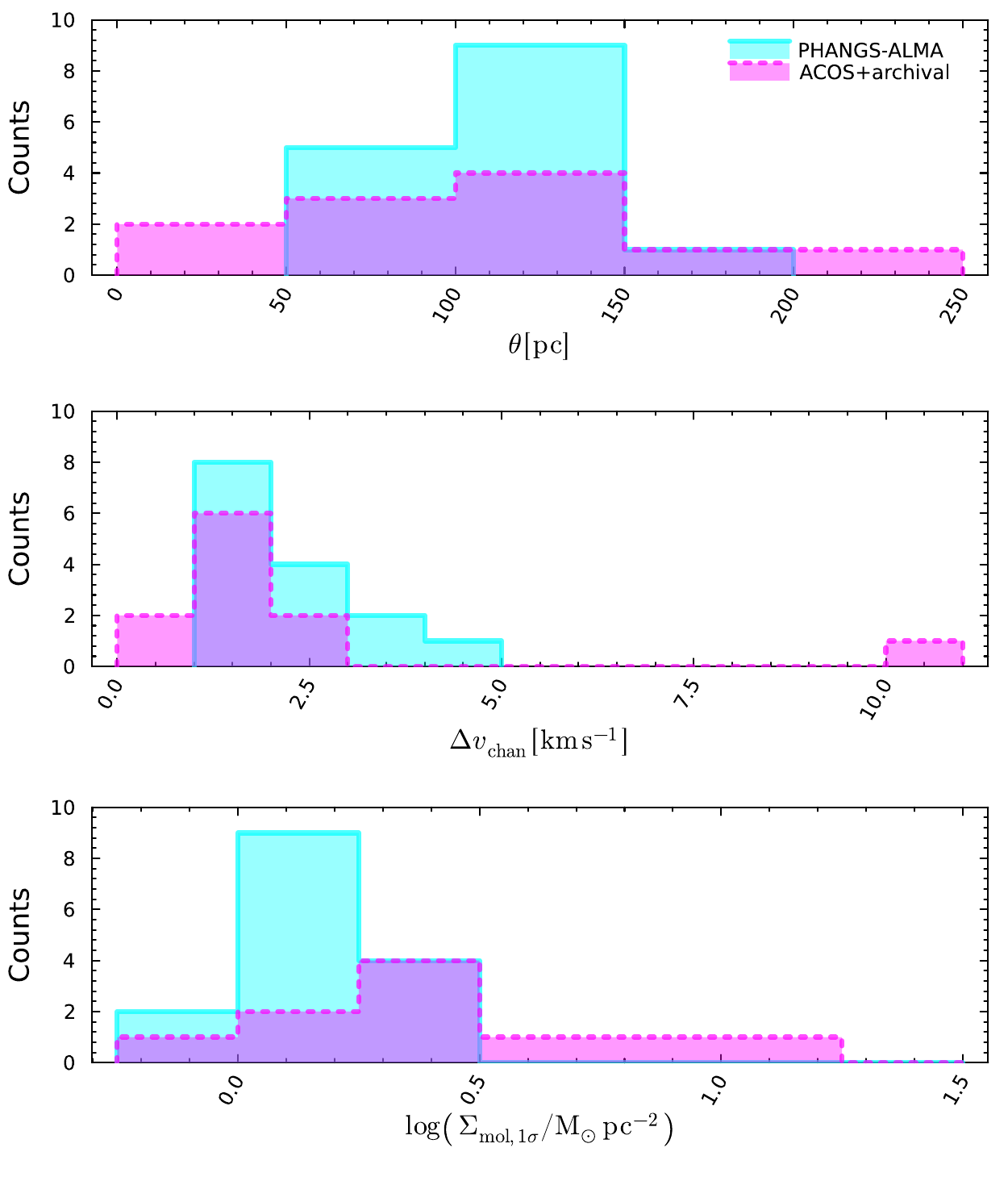}}
    \caption{Histogram comparing ALMA dataset properties from host galaxies of the PHANGS-ALMA sample (solid cyan) and ACOS+archival data (dotted magenta).
    The values are the spatial resolution (top panel), velocity channel width (middle panel), and mass sensitivity (bottom panel).}
    \label{fig:figB1_2}
\end{figure*}

Also from Table~\ref{tab:tabA3}, the characteristics of the cube dataset are included, such as the spatial resolution, velocity resolution, and 1$\sigma$ mass surface density sensitivity.
There is no clear difference for $\theta$ between PHANGS-ALMA and ACOS+archival, except for \object{NGC5128}.
This is also confirmed given the high p-values of the KS and AD two-sample tests as 0.217 and 0.286, respectively.
For $\Delta v_{\rm{chan}}$, the only galaxy with a clear high value is \object{NGC5231}.
However, the low p-values (0.023 for KS and 0.018 for AD) suggest that the samples of PHANGS+ALMA and ACOS+archival belong to different distributions.
The $1\sigma$ mass surface density shows higher values for NGC1614, NGC4038, and \object{PGC084885} than the rest of the data.
When comparing both datasets of PHANGS-ALMA and ACOS+archival, p-values for KS and AD are 0.059 and 0.012, respectively.
Therefore, it is not possible to confirm whether both distributions are similar or not.
% We consider that get rid each of the datasets with higher values than the average would reduced significantly the sample as the essential criteria is to reach cloud scales.
% It is not a significant reduction to remove a few targets. Instead of this say that if we removed them the result will remain unchanged.
If some targets are removed, then the results will remain unchanged.
In all cases, only one parameter is considered as not typical but not for the remaining two parameters.
We notice that the results may differ as both samples (PHANGS-ALMA and ACOS+archival) are slightly different in terms of the features dependent on the ALMA datasets, but the physical interpretation should be similar since the selection effect is not statistically significant.

\begin{landscape}
\begin{table*}[ht!]
    \caption{Host galaxy  properties}
    \label{tab:tabA3}
    \centering

        \begin{tabular}{cccccccccccc}
        \hline
        \hline
        Host galaxy & Dataset & $M_{\rm{K}}$ & Galaxy & $\theta$ & $\Delta v_{\rm{chan}}$ & $\Sigma_{\rm{mol, 1\sigma}}$ & $D_{L}$ & $\log M_{*}$ & $\log SFR$ & {$i$} & \# of\\
        {} & {} & {} & morphology & {} & {} & {} & {} & {} & {} & {} & CCSNe\\
        {} & {} & (mag) & {} & (pc) & (km\,s$^{-1}$) & (M$_{\odot}\,$pc$^{-2}$) & (Mpc) & $ (M_{*}/M_{\odot})$ & $(\mathrm{SFR}/M_{\odot}\,\mathrm{yr}^{-1})$ & {(\degr)} & {}\\
        \hline
        ESO492-002\tablefootmark{a} & ACOS & -23.6 & 3.1 & 94 & 1.3 & 3.1 & 30.0 & 10.08 & 0.4 & 48.8 & 1\\
        NGC0908\tablefootmark{a} & ACOS & -24.4 & 5.1 & 49 & 1.3 & 2.1 & 19.0 & 10.59 & 0.51 & 64.7 & 1\\
        NGC1087\tablefootmark{b,c,d} & PHANGS-ALMA & -23.1 & 5.2 & 123 & 1.4 & 1.2 & 14.4 & 9.81 & 0.01 & 54.1 & 1\\
        NGC1097\tablefootmark{e} & PHANGS-ALMA & -25.0 & 3.3 & 112 & 2.0 & 1.2 & 14.2 & 10.85 & 0.52 & 54.8 & 1\\
        NGC1365\tablefootmark{f} & PHANGS-ALMA & -25.5 & 3.2 & 131 & 2.1 & 1.7 & 18.0 & 10.97 & 0.86 & 62.6 & 1\\
        NGC1559\tablefootmark{e} & PHANGS-ALMA & -23.4 & 5.9 & 118 & 1.4 & 1.5 & 12.6 & 9.91 & 0.2 & 59.9 & 1\\
        NGC1566\tablefootmark{b,c,d} & PHANGS-ALMA & -24.8 & 4.0 & 108 & 1.7 & 1.9 & 18.0 & 10.79 & 0.63 & 49.1 & 1\\
        NGC1614\tablefootmark{g} & archival & -24.7 & 5.0 & 55 & 2.6 & 37.4 & 66.5 & 10.7 & 2.0 & 41.76 & 1\\
        NGC1637\tablefootmark{e} & PHANGS-ALMA & -22.4 & 5.0 & 79 & 2.4 & 0.9 & 9.8 & 9.76 & -0.74 & 31.1 & 1\\
        NGC1672\tablefootmark{h} & PHANGS-ALMA & -24.4 & 3.3 & 182 & 1.4 & 1.5 & 11.9 & 10.36 & 0.3 & 28.9 & 1\\
        NGC2997\tablefootmark{e} & PHANGS-ALMA & -24.5 & 5.1 & 120 & 2.3 & 0.9 & 11.3 & 10.57 & 0.51 & 53.7 & 1\\
        NGC3278\tablefootmark{a} & ACOS & -23.9 & 5.1 & 92 & 1.3 & 4.3 & 42.7 & 10.4 & 0.65 & 41.0 & 1\\
        NGC3354\tablefootmark{a} & ACOS & -22.5 & 5.0 & 107 & 1.3 & 1.6 & 29.8 & 9.55 & -0.32 & 37.3 & 1\\
        NGC3627\tablefootmark{h} & PHANGS-ALMA & -24.2 & 3.1 & 89 & 3.5 & 2.1 & 8.3 & 10.61 & 0.22 & 67.5 & 3\\
        NGC4038\tablefootmark{i} & archival & -24.2 & 8.9 & 44 & 2.6 & 16.3 & 22.0 & 10.54 & 1.0 & 51.94 & 1\\
        NGC4219\tablefootmark{a} & ACOS & -24.2 & 4.0 & 109 & 1.3 & 2.3 & 24.5 & 10.44 & 0.46 & 74.1 & 1\\
        NGC4254\tablefootmark{h} & PHANGS-ALMA & -25.7 & 5.2 & 113 & 1.1 & 1.1 & 16.8 & 10.56 & 0.72 & 20.1 & 4\\
        NGC4303\tablefootmark{h} & PHANGS-ALMA & -24.9 & 4.0 & 149 & 1.3 & 1.4 & 17.6 & 10.72 & 0.77 & 18.1 & 2\\
        NGC4321\tablefootmark{h} & PHANGS-ALMA & -25.1 & 4.0 & 123 & 1.6 & 1.4 & 15.2 & 10.73 & 0.55 & 24.0 & 1\\
        NGC4981\tablefootmark{j} & archival & -23.4 & 4.0 & 152 & 0.6 & 0.8 & 21.0 & 10.17 & -0.01 & 44.7 & 1\\
        NGC5128\tablefootmark{k} & archival & -25.6 & -2.1 & 229 & 0.6 & 1.1 & 3.6 & 10.98 & 0.14 & 45.3 & 2\\
        NGC5231\tablefootmark{l} & archival & -24.7 & 1.1 & 110 & 10.4 & 7.7 & 93.0 & --- & --- & 53.0 & 1\\
        NGC5236\tablefootmark{f,m,n} & PHANGS-ALMA & -24.7 & 5.0 & 51 & 4.9 & 1.0 & 4.89 & 10.53 & 0.62 & 15.3 & 1\\
        NGC5530\tablefootmark{e} & PHANGS-ALMA & -23.1 & 4.2 & 67 & 1.6 & 2.7 & 11.8 & 10.11 & -0.62 & 66.3 & 1\\
        NGC5967\tablefootmark{a} & ACOS & -24.4 & 5.7 & 105 & 1.3 & 2.6 & 31.2 & 10.27 & 0.17 & 50.3 & 1\\
        NGC6744\tablefootmark{h} & PHANGS-ALMA & -24.4 & 4.0 & 52 & 3.0 & 2.8 & 11.6 & 11.1 & 0.63 & 53.5 & 1\\
        PGC084885\tablefootmark{o} & archival & -25.1 & -1.0 & 98 & 5.2 & 31.4 & 79.8 & --- & --- & 53.3 & 1\\
        \hline
    \end{tabular}
    \tablefoot{
    ALMA IDs.\\
    \tablefoottext{a}{2021.1.00099.S}
    \tablefoottext{b}{2015.1.00925.S}
    \tablefoottext{c}{2017.1.00392.S}
    \tablefoottext{d}{2018.1.01651.S}
    \tablefoottext{e}{2017.1.00886.L}
    \tablefoottext{f}{2013.1.01161.S}
    \tablefoottext{g}{2017.1.00255.S}
    \tablefoottext{h}{2015.1.00956.S}
    \tablefoottext{i}{2018.1.00272.S}
    \tablefoottext{j}{2019.1.01305.S}
    \tablefoottext{k}{2018.A.00062.S}
    \tablefoottext{l}{2018.1.00538.S}
    \tablefoottext{m}{2015.1.00121.S}
    \tablefoottext{n}{2016.1.00386.S}
    \tablefoottext{o}{2015.1.00902.S}
    }

\end{table*}
\end{landscape}

\begin{table*}[ht!]
    \caption{KS and AD tests applied to the PHANGS-ALMA and ACOS+archival galaxy properties.}
    \label{tab:tabA4}
    \centering
    \begin{tabular}{l|ccccc}
        \hline
        \hline
        Two samples test & KS statistic & KS p-value & AD statistic & AD p-value\\
        \hline
        $M_{\rm{K}}$ \,\,\,\,\,\,\,\,\,\,\,\,\,\,\,\,\,\,\,\,\,\,\,\,\,\,\,\,\,\,\,\,\,\,\,\,\,\, $\rightarrow$ PHANGS-ALMA vs. ACOS+archival & 0.23 & 0.86 & 0.56 & 0.72 \\
        Galaxy morphology $\rightarrow$ PHANGS-ALMA vs. ACOS+archival & 0.27 & 0.73 & 0.82 & 0.48 \\
        SFR $\rightarrow$ PHANGS-ALMA vs. ACOS+archival                        & 0.23 & 0.900 & 0.6 & 0.640 \\
        M$_{*}$ $\rightarrow$ PHANGS-ALMA vs. ACOS+archival                    & 0.30 & 0.653 & 0.5 & 0.790 \\
        $\theta$ $\rightarrow$ PHANGS-ALMA vs. ACOS+archival                   & 0.42 & 0.217 & 1.1 & 0.286 \\
        $\Delta v_{\rm{chan}}$ $\rightarrow$ PHANGS-ALMA vs. ACOS+archival     & 0.59 & 0.023 & 3.2 & 0.018 \\
        $\Sigma_{\rm mol,1\sigma}$ $\rightarrow$ PHANGS-ALMA vs. ACOS+archival & 0.53 & 0.059 & 3.5 & 0.012 \\
        \hline
    \end{tabular}
\end{table*}

\subsection{Distance of CCSNe from the galactic center}\label{app:appA2sub2}

Galaxy centers contain denser molecular gas in comparison with their disks \citep{2020Sun}.
This means that the molecular gas densities may be biased toward higher values at central positions.
In Figure~\ref{fig:figB2} we display $\Sigma_{\rm{mol}}$, $\sigma_{\rm{CO}(2-1)}$, and $\alpha_{\rm{vir}}$ as a function of the distance to the galaxy center for our CCSN sample (r$_{\rm{SN}}$) and host galaxies.
Furthermore, Fig.~\ref{fig:figB3} shows the normalized distances from the galactic center \citep[r$_{\rm{SN,norm}}$ for CCSNe;][]{1995Bottinelli}, corrected by half of the isophotal diameter $D_{25}$ taken from HyperLeda.
Medians and 1$\sigma$ confidence intervals of these values are reported in Table~\ref{tab:tabA5}.
In addition, in order to compare the distributions of CCSNe and PHANGS-ALMA galaxies, only pixels with normalized distances lower than 0.3 are included in Fig.~\ref{fig:fig1}.
This demonstrates that our CCSNe are not solely located in the central regions or outskirts of the galaxies.
% Fig B2 and B3 shows that SNe are located in central regions. It needs to be specified what you mean here
% DO you mean they are not concentrated solely in the centres?
Host galaxies with and without AGNs are also shown.
Residuals comparing PHANGS-ALMA pixels and CCSN positions are shown to demonstrate an enhanced radial profile (using the parameters $\Sigma_{\rm{mol}}$, $\sigma_{\rm{CO}(2-1)}$, and $\alpha_{\rm{vir}}$) for the latter.
These results indicate that the explosion sites have significantly higher $\Sigma_{\rm{mol}}$, $\sigma_{\rm{CO}(2-1)}$, and $\alpha_{\rm{vir}}$ than typical regions at similar galactocentric radii (Figs.~\ref{fig:figB2} and~\ref{fig:figB3}).

\begin{figure*}[ht!]
    \centering
    \resizebox{\hsize}{!}{\includegraphics{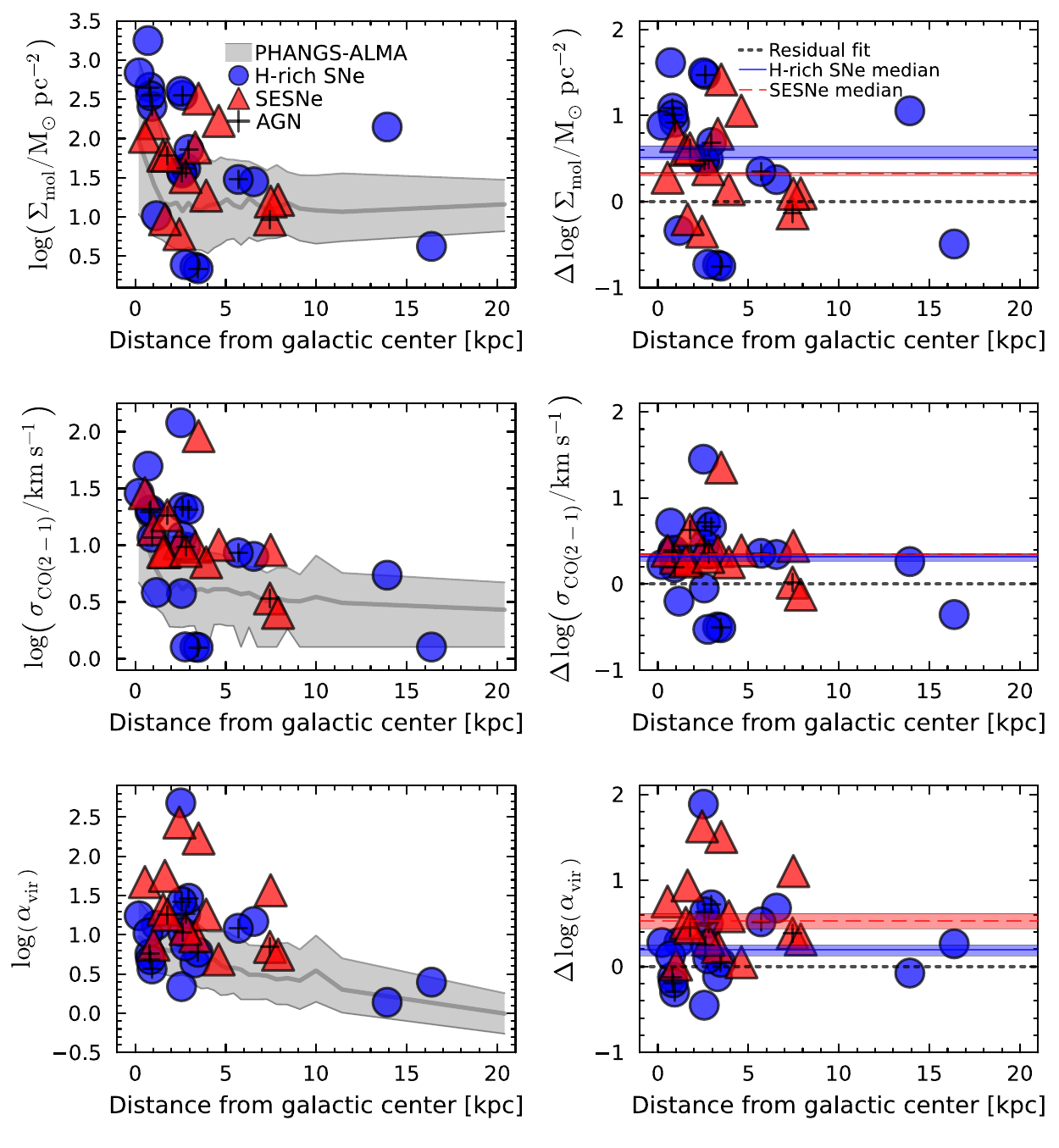}}
    \caption{
    Distances from galactic centers. Molecular gas surface density (top left panel), CO(2--1) velocity dispersion (middle left panel), and virial parameter (bottom left panel) as a function of distance to their respective host galaxy center.
    Blue circles and red triangles represent H-rich SNe and SESNe, respectively.
    Black crosses depict galaxies with an AGN.
    Gray shaded regions show the 16\%, 50\%, and 84\% distribution of host galaxies.
    Residuals with respect to the PHANGS-ALMA pixels, for the molecular gas surface density (top right panel), CO(2--1) velocity dispersion (middle right panel), and virial parameter (bottom right panel) as a function of distance to their respective host galaxy center are shown on the right.
    Median values of residuals are shown as blue solid and red dashed lines (and 1$\sigma$ uncertainties in shaded regions) for H-rich SNe and SESNe, respectively. PHANGS-ALMA residual at zero as a black dotted line.
    }
    \label{fig:figB2}
\end{figure*}

\begin{figure*}[ht!]
    \centering
    \resizebox{\hsize}{!}{\includegraphics{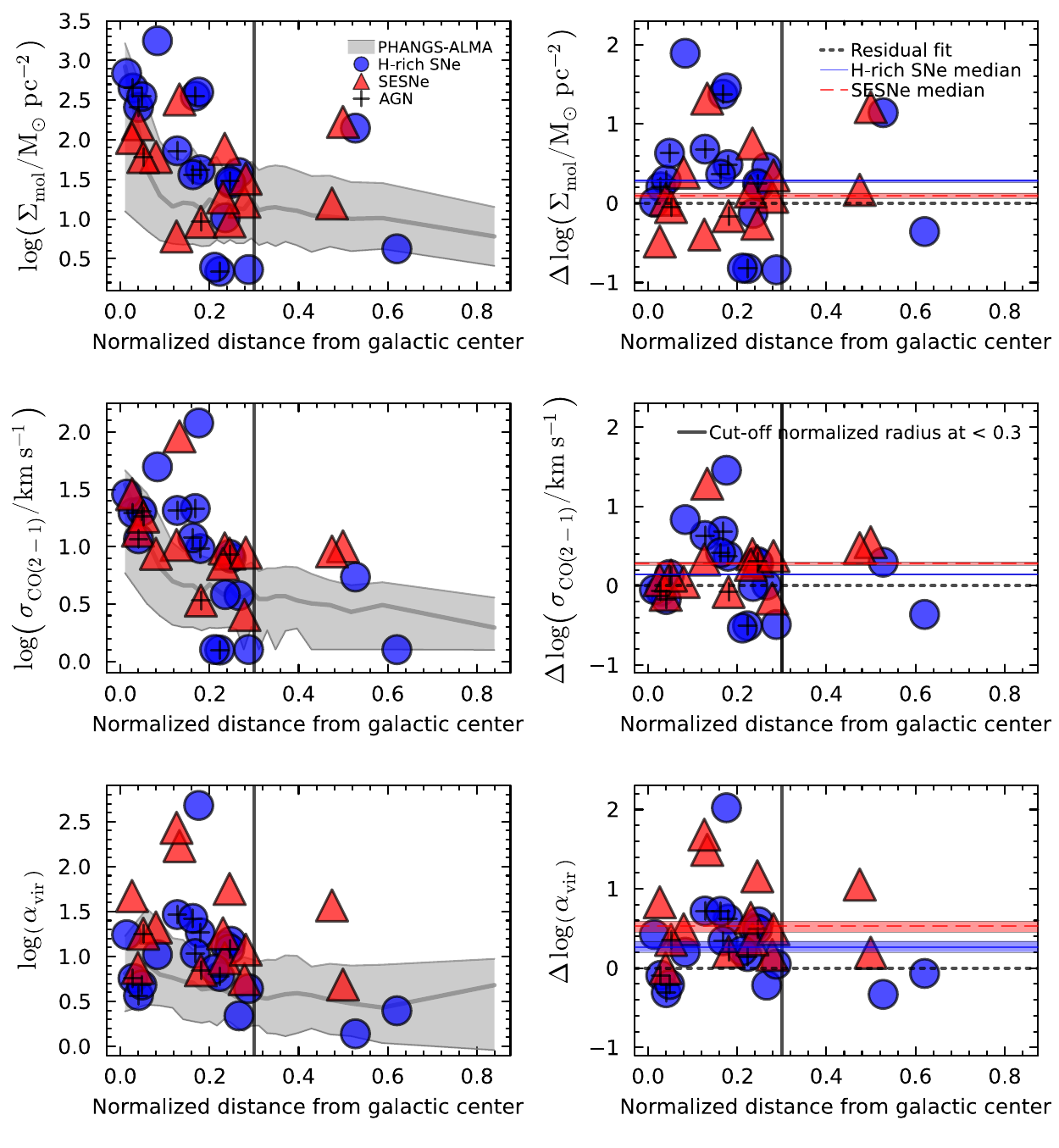}}
    \caption{Similar to Fig.~\ref{fig:figB2} but for normalized galactocentric distances.
    }
    \label{fig:figB3}
\end{figure*}

\begin{table*}[ht!]
    \caption{Medians and 1$\sigma$ confidence intervals for residuals of SN distance from galactic centers}
    \label{tab:tabA5}
    % \centering
    \begin{adjustbox}{width=\columnwidth,center}
    \begin{tabular}{l|cccccc}
        \hline
        \hline
        Distance measurement & \multicolumn{3}{c|}{Distance from galactic center} & \multicolumn{3}{c}{Normalized distance from galactic center}\\
        \hline
        Parameter & $\Delta \log \left( \Sigma_{\rm{mol}} / \rm{M}_{\odot} \, \rm{pc}^{-2} \right)$ & $\Delta \log \left( \sigma_{\rm{CO(2-1)}}  / \rm{km} \, \rm{s}^{-1} \right)$ & $\Delta \log \left( \alpha_{\rm{vir}} \right)$ & $\Delta \log \left( \Sigma_{\rm{mol}} / \rm{M}_{\odot} \, \rm{pc}^{-2} \right)$ & $\Delta \log \left( \sigma_{\rm{CO(2-1)}} / \rm{km} \, \rm{s}^{-1} \right)$ & $\Delta \log \left( \alpha_{\rm{vir}} \right)$\\
        \hline
        H-rich SNe           & $0.51^{+0.12}_{-0.02}$ & $0.32^{+0.02}_{-0.05}$ & $0.19^{+0.06}_{-0.07}$ & $0.24^{+0.01}_{-0.03}$ & $0.14^{+0.01}_{-0.01}$ & $0.27^{+0.07}_{-0.07}$\\
        SESNe                & $0.33^{+0.02}_{-0.02}$ & $0.34^{+0.01}_{-0.02}$ & $0.53^{+0.09}_{-0.09}$ & $0.10^{+0.03}_{-0.03}$ & $0.28^{+0.02}_{-0.02}$ & $0.54^{+0.06}_{-0.06}$\\
        \hline
    \end{tabular}
    \end{adjustbox}
\end{table*}

\subsection{Excess of $\Sigma_{\rm mol}$--$\sigma_{\rm CO}$ and $\Sigma_{\rm mol}$--$\alpha_{\rm vir}$ planes}\label{app:appA2sub3}

To check the enhancement of local gas density at CCSN positions, the excess of $\sigma_{\rm CO(2-1)}$ and $\alpha_{\rm vir}$ with respect to $\Sigma_{\rm mol}$ is computed.
Figure~\ref{fig:figB4} shows $\sigma_{\rm CO(2-1), \, Excess}$ and $\alpha_{\rm vir, \, Excess}$ as a function of $\Sigma_{\rm mol}$ for the CCSNe in our sample.
Medians and 1$\sigma$ confidence intervals of excess values are summarised in Table~\ref{tab:tabA6}.

\begin{figure*}[ht!]
    \centering
    \resizebox{\hsize}{!}{\includegraphics{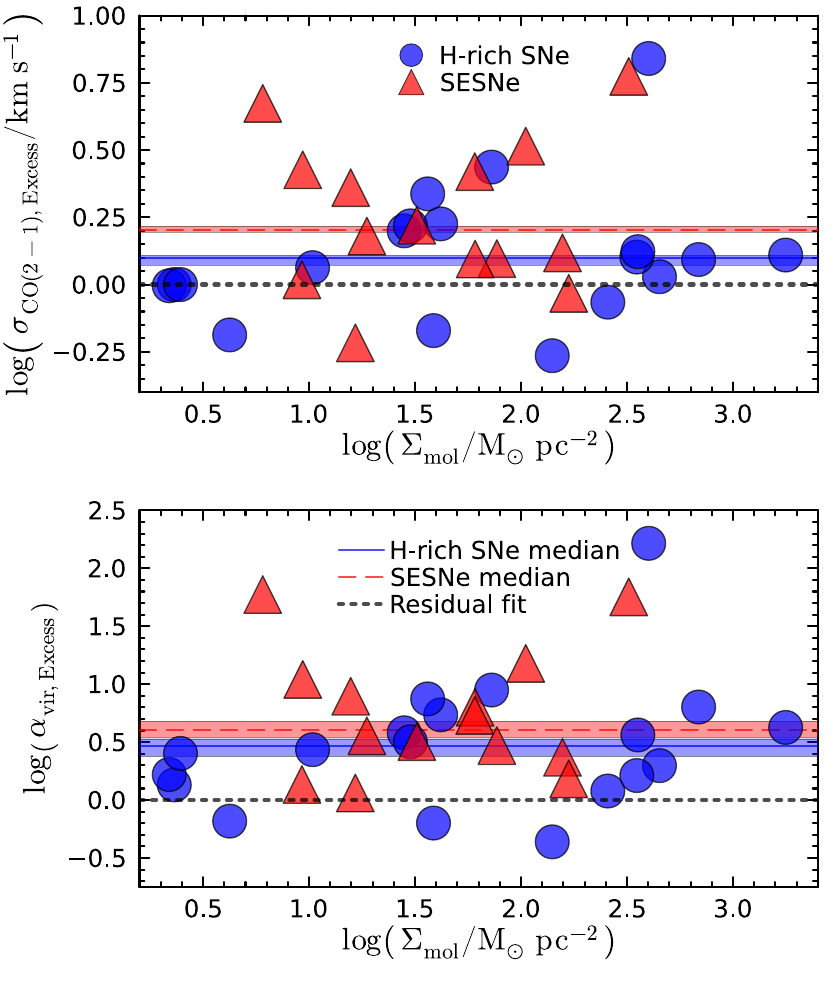}}
    \caption{Excess of $\Sigma_{\rm{mol}}$--$\sigma_{\rm{CO(2-1)}}$ (top panel) and $\Sigma_{\rm{mol}}$--$\alpha_{\rm{vir}}$ (bottom panel) planes for CCSNe.
      H-rich SNe and SESNe represented as blue circles and red triangles, respectively.
      Median values of residuals are shown as blue solid and red dashed lines (and 1$\sigma$ uncertainties in shaded regions) for H-rich SNe and SESNe, respectively.
      PHANGS-ALMA residual at zero is shown as a black dotted line.
    }
    \label{fig:figB4}
\end{figure*}

\begin{table*}[ht!]
    \caption{Medians and 1$\sigma$ confidence intervals for excess of $\sigma_{\rm CO(2-1)}$ and $\alpha_{\rm vir}$ with respect to $\Sigma_{\rm mol}$}
    \label{tab:tabA6}
    \centering
    \begin{tabular}{l|cc}
        \hline
        \hline
        Parameter & $\log \left( \sigma_{\rm{CO(2-1)}, \, Excess} / \rm{km} \, \rm{s}^{-1} \right) $ & $\log \left( \alpha_{\rm{vir, \, Excess}} \right)$\\
        \hline
        H-rich SNe               & $0.11^{+0.01}_{-0.03}$    & $0.47^{+0.06}_{-0.10}$ \\
        SESNe                    & $0.20^{+0.02}_{-0.01}$    & $0.61^{+0.08}_{-0.07}$ \\
        \hline
    \end{tabular}
\end{table*}

\subsection{Host galaxy morphology dependence}\label{app:appA2sub4}

One focus is to determine whether there is a trend in the CCSN explosion sites with respect to the galactic structure in which they occur.
PHANGS-ALMA galaxies identify galactic structures \citep{2021Querejeta}
and basically it classifies ``simple'' galactic regions based on near-infrared Spitzer $3.6~\rm{\mu m}$ IRAC images \citep{2004Fazio}.
This technique has the advantage that stellar light is minimally affected by dust extinction.
Morphological masks are defined as center, bar, disc, interarm, and spiral arm.
We compare each CCSN location with respect to all pixels of $\Sigma_{\rm mol}$--$\sigma_{\rm CO}$ within the respective morphological mask of the host galaxy.
NGC5236 (SN1968L) is not included as the mask is not available.
Also, ACOS and archival data do not have morphological masks and they are not taken into account.
The sample used for this analysis consist of 19 CCSNe (14 H-rich SNe and 5 SESNe), from which there are three in the center, one in the bar, three in the disc, seven in the interarm, and five in the spiral arm.

Figures~\ref{fig:figB6_1},~\ref{fig:figB6_2}, and~\ref{fig:figB6_3} show the relation of $\sigma_{\rm{CO(2-1)}}$ as a function of $\Sigma_{\rm{mol}}$ for the CCSN sample used and the respective galactic structures in the host galaxies.
It is worth noting that for two host galaxies, each contain two CCSNe (SN1973R and SN2009hd in NGC3627; SN1972Q and SN1986l in NGC4254) located within the same galactic structure.
Fig.~\ref{fig:figB6_2} includes the CCSNe in the same panel for those cases.
Similarly as in Section~\ref{app:appA2sub3}, the excess of $\Sigma_{\rm mol}$--$\sigma_{\rm CO}$ are computed, resulting in the 16th, 50th, and 84th percentiles of the distribution of -0.142, 0.002, and 0.308 dex, respectively.
% The general trends show consistent CCSN environments compared to their galactic structures. 
More precisely, the general trends show that CCSN molecular gas environments share similar properties of the galactic structures in which they occur.
However, this is based on a reduced sample of 19 CCSNe ($\sim70\%$ of them being H-rich SNe; only PHANGS-ALMA data).

\begin{figure*}[ht!]
    \centering
    \resizebox{\hsize}{!}{\includegraphics{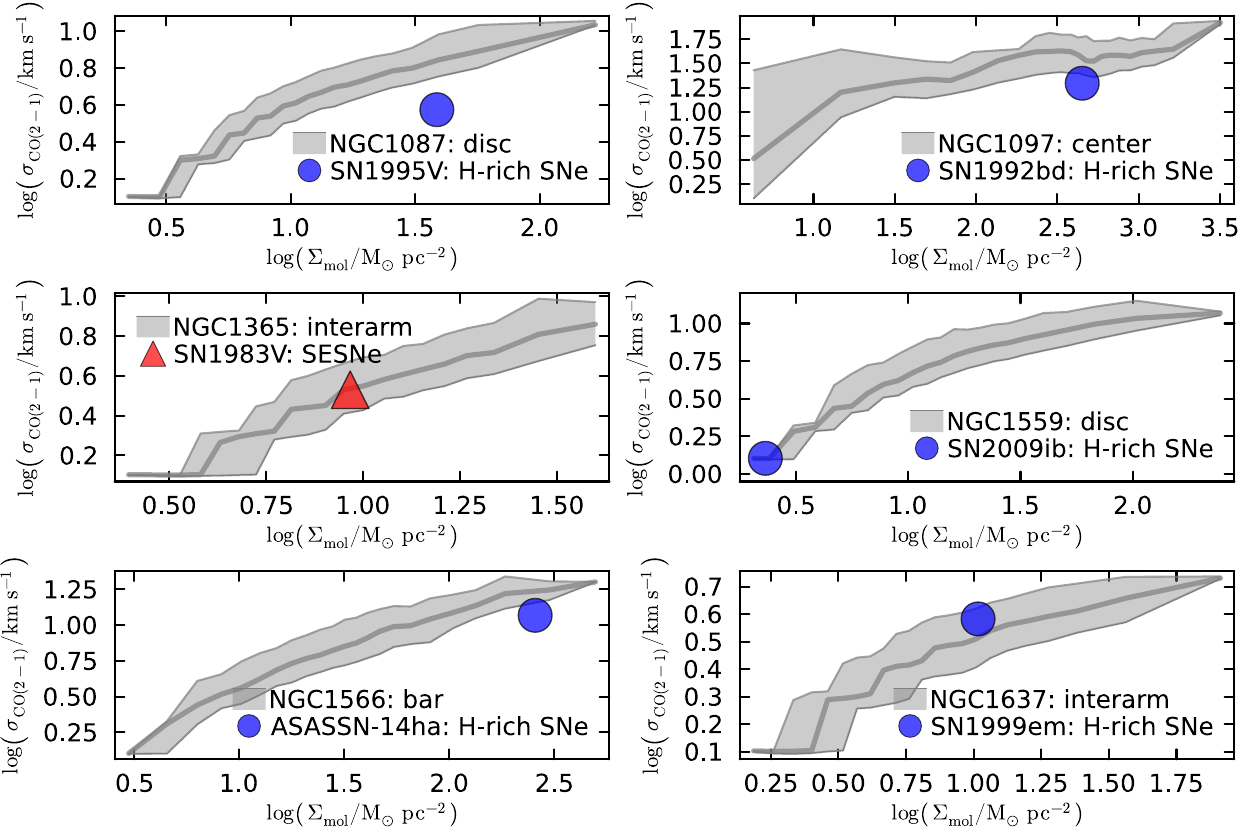}}
    \caption{CCSN environment and host galaxy morphology relations for $\sigma_{\rm{CO}(2-1)}$ as a function of $\Sigma_{\rm{mol}}$.
    Locations of H-rich SNe, SESNe, and PHANGS-ALMA galaxy pixels are represented as blue circles, red triangles, and gray shaded regions, respectively.
    }
    \label{fig:figB6_1}
\end{figure*}

\begin{figure*}[ht!]
    \centering
    \resizebox{\hsize}{!}{\includegraphics{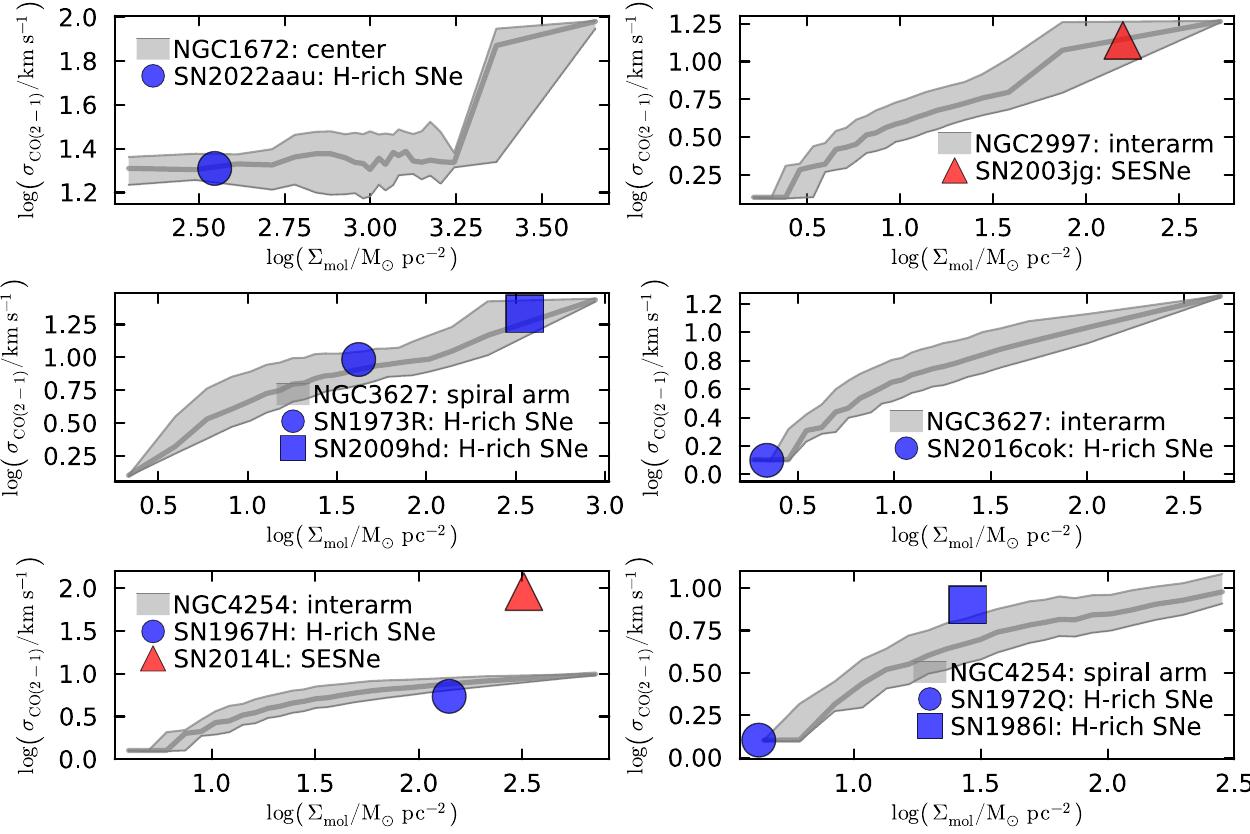}}
    \caption{Continuation of Fig.~\ref{fig:figB6_1} but blue squares also represent H-rich SNe.
    }
    \label{fig:figB6_2}
\end{figure*}

\begin{figure*}[ht!]
    \centering
    \resizebox{\hsize}{!}{\includegraphics{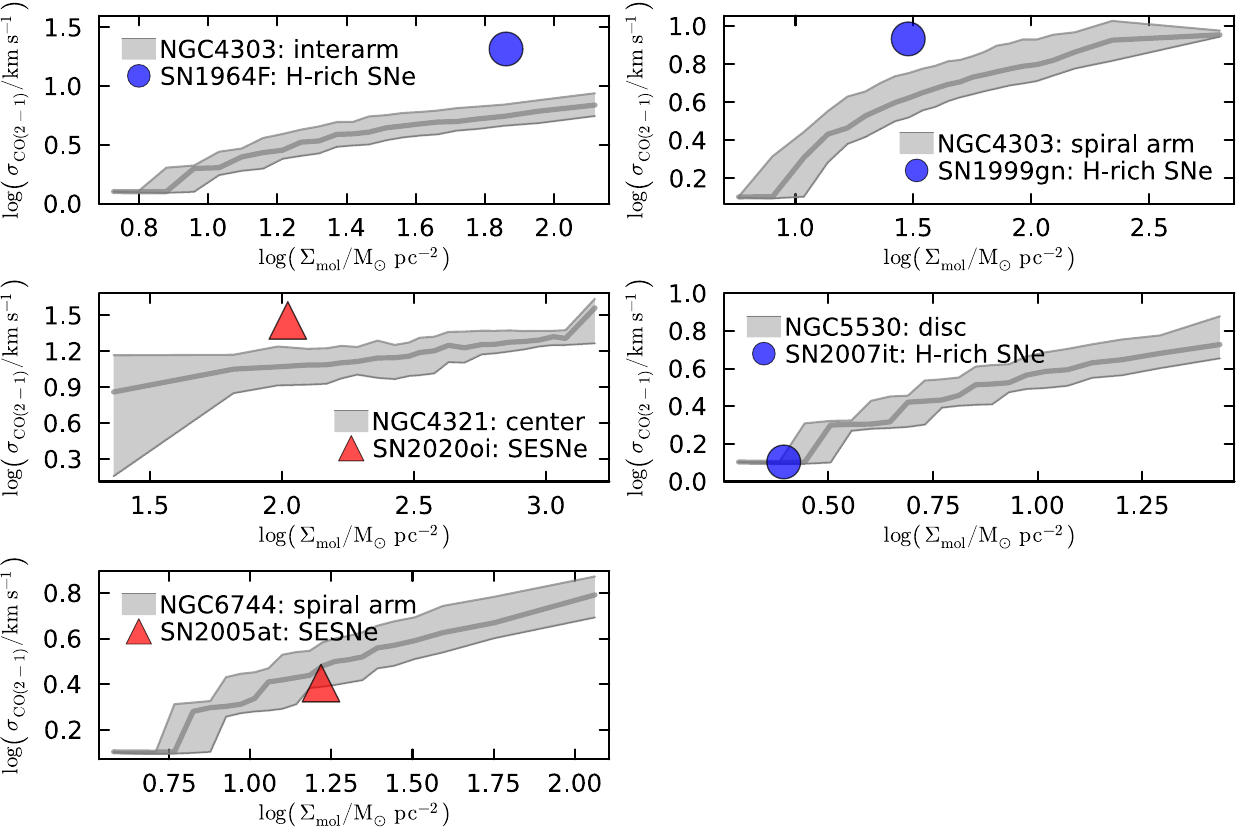}}
    \caption{Continuation of Fig.~\ref{fig:figB6_1} and Fig.~\ref{fig:figB6_2}.}
    \label{fig:figB6_3}
\end{figure*}

\end{appendix}
\end{document}